# Capacity-Achieving Ensembles of Accumulate-Repeat-Accumulate Codes for the Erasure Channel with Bounded Complexity*

Henry D. Pfister
EPFL
Lausanne 1015, Switzerland
henry.pfister@epfl.ch

Igal Sason
Technion
Haifa 32000, Israel
sason@ee.technion.ac.il

August 6, 2018

**Abstract**

The paper introduces ensembles of accumulate-repeat-accumulate (ARA) codes which asymptotically achieve capacity on the binary erasure channel (BEC) with *bounded complexity*, per information bit, of encoding and decoding. It also introduces symmetry properties which play a central role in the construction of capacity-achieving ensembles for the BEC with bounded complexity. The results here improve on the tradeoff between performance and complexity provided by previous constructions of capacity-achieving ensembles of codes defined on graphs. The superiority of ARA codes with moderate to large block length is exemplified by computer simulations which compare their performance with those of previously reported capacity-achieving ensembles of LDPC and IRA codes. The ARA codes also have the advantage of being systematic.

# 1 Introduction

Error-correcting codes which employ iterative decoding algorithms are now considered state of the art in the field of low-complexity coding techniques. By now, there is a large collection of families of iteratively decoded codes including low-density parity-check (LDPC), turbo, repeat-accumulate and product codes; all of them demonstrate a rather small gap (in rate) to capacity with feasible complexity.

The study of capacity-achieving (c.a.) sequences of LDPC codes for the binary erasure channel (BEC) was initiated by Luby et al. [1] and Shokrollahi [2]. They show that it is possible to closely approach the capacity of an erasure channel with a simple iterative procedure whose complexity is linear in the block length of the code [1, 2]. Following these works, Oswald and Shokrollahi presented in [3] a systematic study of c.a. sequences of LDPC codes for the BEC. Jin et al. introduced irregular repeat-accumulate (IRA) codes and presented a c.a. sequence of systematic IRA (SIRA) codes for the BEC [4]. A sequence of c.a. SIRA codes for the BEC with lower encoding and decoding complexities was introduced in [5, Theorem 2]. All of the aforementioned codes have one drawback in common: their decoding complexity scales like the log of the inverse of the gap (in rate) to capacity [2, 3, 5, 6, 7]; hence, under iterative message-passing decoding, these codes have *unbounded complexity* (per information bit) as the gap to capacity vanishes.

In [8], the authors presented for the first time two sequences of ensembles of non-systematic IRA (NSIRA) codes which asymptotically (i.e., as their block length tends to infinity) achieve capacity on the BEC with *bounded complexity* per information bit. This new result is achieved by puncturing bits and thereby introducing state nodes in the Tanner graph representing the codes. We note that for fixed complexity, these codes will eventually (for large enough block length) outperform any code proposed so far. However, the speed of convergence happens to be quite slow and, for small to moderate block lengths, the codes introduced in [8] are not record breaking.

In this paper, we are interested in the construction and analysis of c.a. codes for the BEC with bounded complexity that also perform well at moderate block lengths. We also would also like these codes to be systematic and to have reasonably low error floors. To this end, we make use of a new channel coding scheme, called "Accumulate-Repeat-Accumulate" (ARA) codes, which was recently introduced by Abbasfar et al. [9]. These codes are systematic and have both outstanding performance, as exemplified in [9, 10, 11], and a simple linear-time encoding. After presenting an appropriate ensemble of irregular ARA codes, we construct a number of c.a. degree distributions. Simulations show that some of these ensembles perform quite well on the BEC at moderate block lengths. We therefore expect that irregular ARA codes, optimized for general channels, also perform well at moderate block lengths (as is partially supported by some simulation results in [9]). This issue is regarded as a topic for further research, while this paper is focused on the BEC. Throughout the paper, we consider the encoding and decoding complexity *per information bit.*

Along the way, we study symmetry properties of c.a. sequences for the BEC and discover a new code structure which we call "Accumulate-LDPC" (ALDPC) codes. We show that c.a. degree distributions for this structure can be easily constructed based on the results of [8, Theorems 1, 2]. This fact and structure was proposed independently by Hsu and Anastasopoulos [12].

The paper is organized as follows: Section 2 introduces ARA codes, describes their encoding and decoding, and their density evolution analysis for the BEC. Section 3 introduces symmetry properties that play a central role in the construction of c.a. sequences of ensembles for the BEC. Section 4 serves as a preparatory step towards the construction of explicit c.a. sequences of ARA codes for the BEC, where their complexity of encoding and decoding stays bounded as the gap to capacity vanishes. Section 5 presents explicit constructions of c.a. sequences of bit-regular and check-regular ARA codes with bounded complexity. Section 6 focuses on the construction of c.a. ensembles of ARA, NSIRA and ALDPC codes (with bounded complexity) based on the ensembles of self-matched LDPC codes introduced in the section. Computer simulations for the BEC are presented in Section 7, and the superiority of self-matched ARA codes with moderate to large block length is exemplified by comparing their performance with those of previously reported c.a. ensembles of LDPC and IRA codes from [2, 8]. Finally, Section 8 concludes our discussion.

# 2 Accumulate-Repeat-Accumulate Codes

In this section, we present our ensemble of ARA codes. Density evolution (DE) analysis of this ensemble is presented in the second part of this section using two different approaches which lead to the same "DE fixed point equation"; this equation characterizes the fixed points of the iterative message-passing decoder. The connection between these two approaches is used later in this paper to state some symmetry properties which serve as an analytical tool for designing various c.a. ensembles for the BEC (e.g., ARA, IRA and ALDPC codes).

## 2.1 Description of ARA Codes

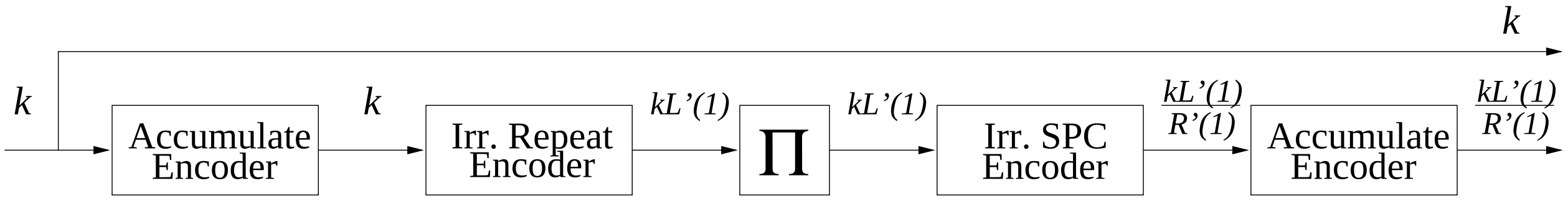

Figure 1: Block diagram for the systematic ARA ensemble ('Irr.' and 'SPC' stand for 'irregular' and 'single-parity check', respectively, and Π stands for a bit interleaver.)

ARA codes can be viewed either as interleaved serially concatenated codes (i.e., turbo-like codes) or as sparse-graph codes (i.e., LDPC-like codes). From an encoding point of view, it is more natural to view them as interleaved serially concatenated codes (see Fig. 1) where the encoding process is described in Section 2.2.

Since the decoding algorithm of ARA codes is simply belief propagation on the appropriate Tanner graph (see Fig. 2), this leads one to view them as sparse-graph codes from a decoding point of view. Treating these codes as sparse-graph codes also allows one to build large codes by "twisting" together many copies of a single small *protograph* [13, 14]. In general, this approach leads to very good codes with computationally efficient decoders.

In this work, we consider the ensemble of irregular ARA codes which is the natural generalization of the IRA codes from [4]. The ensemble of irregular ARA codes differs slightly from those proposed in [9, 10, 11]. For this ensemble, we find that DE for the BEC can be computed in closed form and that algebraic methods can be used to construct c.a. sequences.

## 2.2 Encoding of ARA Codes

We describe here briefly the encoding process of the ARA codes in Fig. 1. The encoding of ARA codes is done as follows: first, the information bits are accumulated (i.e., differentially encoded), and then the bits are repeated a varying number of times (by an irregular repetition code) and interleaved. The interleaved bits are partitioned into disjoint sets (whose size is not fixed in general), and the parity of each set of bits is computed. Finally, the bits are accumulated for the second time. A codeword of systematic ARA codes is composed of the information bits and the parity bits at the output of the second accumulator.

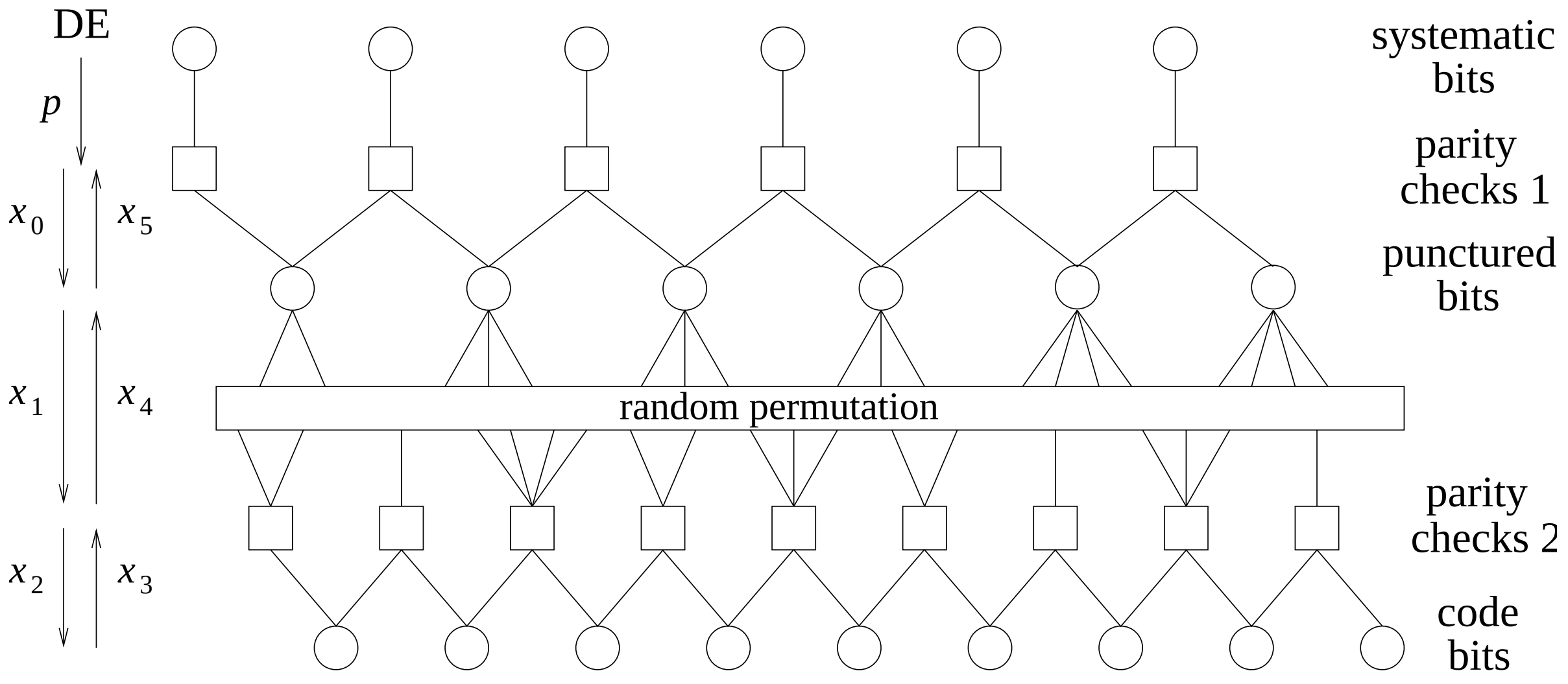

Figure 2: Tanner graph for the ARA ensemble.

Some slight modifications are used later for our simulations and these details are explained in Section 7. In this section, all references to the decoding graph should be taken to imply Fig. 2, and all sums are assumed to be modulo-2.

We will refer to the three layers of bit nodes in the decoding graph as systematic bits, punctured bits, and parity bits (the parity bits are named as "code bits" in Fig. 2). Referring to the Tanner graph of ARA codes, we designate the systematic bits from left to right by $(u_1, u_2, \ldots, u_k)$. The same convention is used for the punctured bits $(v_1, v_2, \ldots, v_k)$ and the parity bits $(z_1, z_2, \ldots, z_{n-k})$.

From the upper part of the graph, it follows that $v_j = u_j \oplus v_{j-1}$ for $j \in \{2, \ldots, k\}$ and $v_1 = u_1$. This yields that

$$v_j = \sum_{i=1}^{j} u_i \qquad j = 1, 2, \ldots, k. \tag{1}$$

Let $d(i)$ be the degree of the $i$-th "parity-check 2" node where the degree is w.r.t. the edges connecting the "punctured bit" nodes and the "parity-check 2" nodes, and $c(i, j)$ be the index of the punctured bit attached to the $j$-th edge of the $i$-th "parity-check 2" node. All the connections between the "punctured bit" nodes and "parity-check 2" nodes are described by these two sequences. Let the sequence $(w_1, w_2, \ldots, w_{n-k})$ be defined by

$$w_i \triangleq \sum_{j=1}^{d(i)} v_{c(i,j)} \qquad i = 1, 2, \ldots, n-k.$$

This can be thought as the sum of the punctured bits which are connected to the $i$-th "parity-check 2" node. From the lower part of the graph we have $z_j = z_{j-1} \oplus w_j$ (where $w_0 \triangleq 0$), and this

gives

$$z_j = \sum_{i=1}^{j} w_i \qquad j = 1, 2, \dots, n-k. \tag{2}$$

From Fig. 2 and equations (1) and (2), one can see that an ARA code is the serial concatenation of four simple codes. The first is an accumulate code (upper part of the graph), the second is an interleaved irregular repetition code, the third is an irregular single parity-check (SPC) code (which is an irregular code due to the varying degrees of the "parity-check 2" nodes), and finally the fourth is a second accumulate code (lower part of the graph).

## 2.3 Density Evolution of Systematic ARA Ensembles

We consider here the asymptotic analysis of ensembles of ARA codes where it is assumed that the codes are transmitted over a BEC and decoded with an iterative message-passing decoder. Based on the density evolution (DE) equations, derived in terms of the degree distributions of these ensembles, we consider the fixed points of the decoding process. In the following, we present two different approaches for the DE analysis of ARA codes for the BEC which, as expected, provide equivalent results. While the concept of the first approach is standard, the second one is helpful in establishing symmetry properties of c.a. ensembles for the BEC; these symmetries are discussed later in Section 3.

### 2.3.1 Density Evolution via Message Passing

An irregular ensemble of ARA codes is defined by its degree distribution (d.d.). Nodes in the decoding graph will be referred to by the names given in Fig. 2. Let $L(x) = \sum_{i=1}^{\infty} L_i x^i$ be a power series where $L_i$ denotes the fraction of "punctured bit" nodes with degree-$i$. Similarly, let $R(x) = \sum_{i=1}^{\infty} R_i x^i$ be a power series where $R_i$ denotes the fraction of "parity-check 2" nodes with degree-$i$. In both cases, the degree refers only to the edges connecting the "punctured bit" nodes to the "parity-check 2" nodes. Similarly, let $\lambda(x) = \sum_{i=1}^{\infty} \lambda_i x^{i-1}$ and $\rho(x) = \sum_{i=1}^{\infty} \rho_i x^{i-1}$ form the d.d. pair from the edge perspective where $\lambda_i$ and $\rho_i$ designate the fraction of the edges which are connected to "punctured bit" nodes and "parity-check 2" nodes with degree-$i$, respectively. We also assume that the permutation in Fig. 1 is chosen uniformly at random from the set of all permutations. The pair of degree distributions of an ARA ensemble is given by $(\lambda, \rho)$.

It is easy to show the following connections between the d.d. pairs w.r.t. the nodes and the edges in the graph:

$$\lambda(x) = \frac{L'(x)}{L'(1)}\,, \qquad \rho(x) = \frac{R'(x)}{R'(1)} \tag{3}$$

or equivalently, since $L(0) = R(0) = 0$, then

$$L(x) = \frac{\int_0^x \lambda(t)\,dt}{\int_0^1 \lambda(t)\,dt}\,, \qquad R(x) = \frac{\int_0^x \rho(t)\,dt}{\int_0^1 \rho(t)\,dt}\,. \tag{4}$$

The design rate $R$ of the ensemble of ARA codes (see Fig. 1) is computed by expressing the block length $n$ as the sum of $k$ systematic bits and $kL'(1)/R'(1)$ parity bits which then yields

$$R = \frac{1}{1 + \frac{L'(1)}{R'(1)}}. \tag{5}$$

A random code is chosen from the ensemble and a random codeword is transmitted over a BEC with erasure probability $p$. The asymptotic performance of the iterative message-passing decoder (as the block length of the code tends to infinity) is analyzed by tracking the average fraction of erasure messages which are passed in the graph of Fig. 2 during the $l^{\text{th}}$ iteration. The technique was introduced in [15] and is known as density evolution (DE). The main assumption of density evolution is that the messages passed on the edges of the Tanner graph are statistically independent. This assumption is justified by the fact that, for randomly chosen codes, the fraction of bits involved in finite-length cycles vanishes as the block length tends to infinity.

A single decoding iteration consists of six smaller steps which are performed on the Tanner graph of Fig. 2. Messages are first passed downward from the "systematic bit" nodes through each layer to the "code bit" nodes. Then, messages are passed back upwards from the "code bit" nodes through each layer to the "systematic bit" nodes. Let $l$ designate the iteration number. Referring to Fig. 2, let $x_0^{(l)}$ and $x_5^{(l)}$ designate the probabilities of an erasure message from the "parity-check 1" nodes to the "punctured bit" nodes and vice-versa, let $x_1^{(l)}$ and $x_4^{(l)}$ be the probabilities of an erasure message from the "punctured bit" nodes to the "parity-check 2" nodes and vice versa, and finally, let $x_2^{(l)}$ and $x_3^{(l)}$ be the probabilities of an erasure message from the "parity-check 2" nodes to "code bit" nodes and vice versa.

From the Tanner graph of ARA codes in Fig. 2, an outgoing message from a "parity-check 1" node to a "punctured bit" node is an erasure if and only if the incoming message through the other edge which connects a "punctured bit" node to the same "parity-check 1" node is an erasure, and also the message received from the BEC for the systematic bit which is connected to the same "parity-check 1" node is also an erasure. Using the statistical independence assumption, this yields the recursive equation

$$x_0^{(l)} = 1 - (1-p)\left(1 - x_5^{(l-1)}\right).$$

It is also clear from Fig. 2 that an outgoing message from a "punctured bit" node to a "parity-check 2" node is an erasure if and only if all the incoming messages passed through the edges which connect the other "parity-check 2" nodes to the same "punctured bit" node are erasures, and also the incoming messages which are passed through the two edges connecting the "parity-check 1" nodes and the considered "punctured bit" node are erasures. The update rule of the iterative message-passing decoder on the BEC therefore implies that

$$x_1^{(l)} = \left(x_0^{(l)}\right)^2 \lambda\left(x_4^{(l-1)}\right).$$

From the graph in Fig. 2, we obtain in a similar manner the following DE equations of the iterative message-passing decoder:

$$\begin{aligned}
x_0^{(l)} &= 1 - \left(1 - x_5^{(l-1)}\right)(1-p) \\
x_1^{(l)} &= \left(x_0^{(l)}\right)^2 \lambda\left(x_4^{(l-1)}\right) \\
x_2^{(l)} &= 1 - R\left(1 - x_1^{(l)}\right)\left(1 - x_3^{(l-1)}\right) \qquad l = 1, 2, \ldots \\
x_3^{(l)} &= p x_2^{(l)} \\
x_4^{(l)} &= 1 - \left(1 - x_3^{(l)}\right)^2 \rho\left(1 - x_1^{(l)}\right) \\
x_5^{(l)} &= x_0^{(l)} \, L\left(x_4^{(l)}\right)
\end{aligned}$$

A fixed point is implied by

$$\lim_{l \to \infty} x_i^{(l)} \triangleq x_i \qquad i = 0, 1, \ldots, 5.$$

Now, we can solve for the fixed point by substituting $x_2$ into $x_3$, and then substituting the result into $x_4$ which gives the fixed point equation

$$x_4 = 1 - \left( \frac{1-p}{1-p\,R(1-x_1)} \right)^2 \rho(1-x_1). \tag{6}$$

Likewise, putting $x_0$ into $x_5$ gives the fixed point equation

$$x_5 = \frac{p\,L(x_4)}{1-(1-p)\,L(x_4)}$$

and plugging this into $x_0$ gives

$$x_0 = 1-(1-x_5)(1-p) = \frac{p}{1-(1-p)L(x_4)}. \tag{7}$$

Finally, Eqs. (6), (7) and the equality $x_1 = x_0^2 \lambda(x_4)$ give the following implicit equation for $x_1 \triangleq x$:

$$\frac{p^2\,\lambda\left(1-\left(\frac{1-p}{1-pR(1-x)}\right)^2 \rho(1-x)\right)}{\left[1-(1-p)\,L\left(1-\left(\frac{1-p}{1-pR(1-x)}\right)^2 \rho(1-x)\right)\right]^2} = x. \tag{8}$$

This equation provides the fixed points of the iterative message-passing decoder.

### 2.3.2 Density Evolution via Graph Reduction

For ensembles of ARA codes whose transmission takes place over a BEC, the DE fixed point equation (8) can be also derived using a *graph reduction* approach. This approach introduces two new operations on the Tanner graph which remove nodes and edges while preserving the information in the graph.

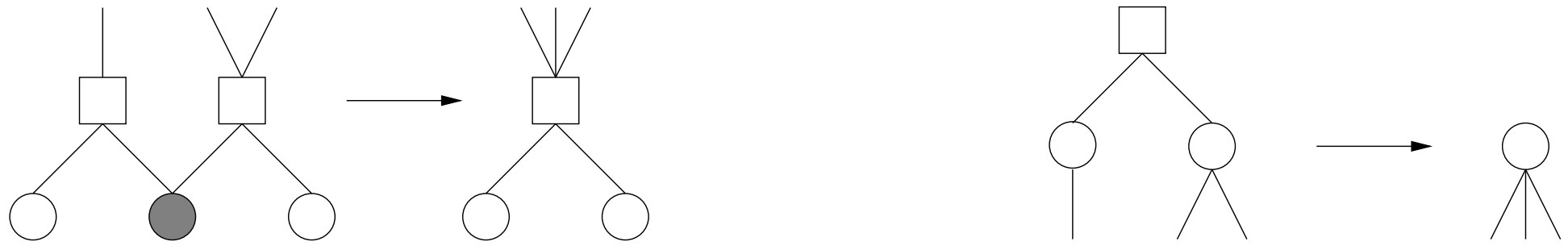

Figure 3: Graph reduction operation applied to parity-check nodes (left) and bit nodes (right).

We start by noting that any "code bit" node whose value is not erased by the BEC can be removed from the graph by absorbing its value into its two "parity-check 2" nodes. On the other hand, when the value of a "code bit" node is erased, one can merge the two "parity-check 2" nodes which are connected to it (by summing the equations) and then remove the "code bit" node from the graph. This merging of two "parity-check 2" nodes causes their degrees to be summed and is shown on the left in Figure 3. Now, we consider the degree distribution (d.d.) of a single "parity-check 2" node in the reduced graph. This can be visualized as working from left to right in the graph, and assuming the value of the previous "code bit" node was known. The probability that there are $k$ erasures before the next observed "code bit" is given by $p^k(1-p)$. The graph reduction associated with this event causes the degrees of $k+1$ "parity-check 2" nodes to be summed. The generating function for this sum of $k+1$ random variables, each chosen independently from the

d.d. $R(x)$, is given by $R(x)^{k+1}$. Therefore, the new d.d. of the "parity-check 2" nodes after the graph reduction is given by

$$\widetilde{R}(x) = \sum_{k=0}^{\infty} p^k (1-p) R(x)^{k+1} = \frac{(1-p)R(x)}{1-pR(x)}. \tag{9}$$

A similar graph reduction can be also performed on the "systematic bit" nodes in Fig. 2. Since degree-1 bit nodes (e.g., the "systematic bit" nodes in Fig. 2) only provide channel information, erasures make them worthless. So they can be removed along with their parity-checks (i.e., the "parity-check 1" nodes in Fig. 2) without affecting the decoder. On the other hand, whenever the value of a "systematic bit" node is observed (assume the value is zero w.o.l.o.g.), it can be removed leaving a degree-2 parity-check. Of course, degree-2 parity-checks imply equality and allow the connected "punctured bit" nodes to be merged (effectively summing their degrees). This operation is shown on the right in Figure 3. The symmetry between graph reduction on the information bits and the parity checks will become important later. Now, we consider the d.d. of a single "punctured bit" node in the reduced graph. This can be seen as working from left to right in the graph, and assuming the value of the previous "systematic bit" node was erased. The probability of the event where the values of $k$ "systematic bit" nodes are observed and the value of the next "systematic bit" node is erased by the channel is given by $(1-p)^k p$. The graph reduction associated with this event causes the degrees of $k+1$ "punctured bit" nodes (from the d.d. $L(x)$) to be summed. Hence, the new d.d. of the "punctured bit" nodes after graph reduction is given by

$$\widetilde{L}(x) = \sum_{k=0}^{\infty} (1-p)^k p L(x)^{k+1} = \frac{pL(x)}{1-(1-p)L(x)}. \tag{10}$$

After the graph reduction, we are left with a standard LDPC code with new edge-perspective degree distributions given by

$$\widetilde{\lambda}(x) = \frac{\widetilde{L}'(x)}{\widetilde{L}'(1)} = \frac{p^2 \lambda(x)}{\big(1-(1-p)L(x)\big)^2} \tag{11}$$

$$\widetilde{\rho}(x) = \frac{\widetilde{R}'(x)}{\widetilde{R}'(1)} = \frac{(1-p)^2 \rho(x)}{\big(1-pR(x)\big)^2}. \tag{12}$$

After the aforementioned graph reduction, all the "systematic bit" nodes and "code bit" nodes are removed. Therefore the residual LDPC code effectively sees a BEC whose erasure probability is 1, and the DE fixed point equation is given by

$$\widetilde{\lambda}\big(1-\widetilde{\rho}(1-x)\big) = x. \tag{13}$$

Based on (11) and (12), the last equation is equivalent to (8).

**Remark 1 (The notation of tilted degree distributions).** The tilted degree distributions $\widetilde{\lambda}$ and $\widetilde{\rho}$ which are given in (11) and (12), respectively, depend on the erasure probability of the BEC ($p$). For simplicity of notation, we do not write this dependency explicitly in our notation. However, in Section 3, when discussing symmetry properties and replacing $p$ by $1-p$, the erasure probability is written explicitly in these tilted degree distributions.

## 2.4 The Stability Condition for ARA Codes

Like the NSIRA codes presented in [8], ARA codes have DE fixed points at both $x=0$ and $x=1$. One can see this by evaluating (8) at these points while assuming that each d.d. function $f$ satisfies $f(0) = 0$ and $f(1) = 1$. To get decoding started, the d.d. is perturbed slightly by adding degree-1 parity-checks, pilot bits, and/or systematic bits. For successful completion of decoding, we need the fixed point at $x = 0$ to be *stable*. To minimize the number of extra bits required to get decoding started, it is also useful for the fixed point (prior to the perturbation) at $x = 1$ to be *unstable*. Although $x = 1$ is not a fixed point after the perturbation, the the *instability* condition helps prevent the decoder from getting stuck near $x = 1$.

The stability and instability conditions are computed by taking the derivative of the LHS of (8) at $x = 0$ and $x = 1$. For the fixed point at $x = 0$ to be stable, we need the derivative to be less than unity, and this gives

$$p^2\lambda_2\left(\rho'(1) + \frac{2pR'(1)}{1-p}\right) < 1. \tag{14}$$

Ensembles without degree-2 bits are unconditionally stable at $x = 0$.

For the fixed point at $x = 1$ to be unstable, we need the derivative to be greater than unity, and this gives

$$(1-p)^2\rho_2\left(\lambda'(1) + \frac{2(1-p)L'(1)}{p}\right) > 1. \tag{15}$$

This condition requires the presence of a non-vanishing fraction of degree-2 "parity-check 2" nodes; ensembles not having this property are unable to immediately create new degree-1 checks and may therefore get stuck shortly after starting. The instability condition guarantees that, on average, more new degree-1 checks are being created than lost when $x$ is close to 1.

# 3 Symmetry Properties of Capacity-Achieving Codes

In this section, we discuss the symmetry between the bit and check degree distributions of c.a. ensembles for the BEC. First, we describe this relationship for LDPC codes, and then we extend it to ARA codes. The extension is based on analyzing the decoding of ARA codes in terms of graph reduction and the DE analysis of LDPC codes.

## 3.1 Symmetry Properties of Capacity-Achieving LDPC Codes

The relationship between the bit d.d. and check d.d. of c.a. ensembles of LDPC codes can be expressed in a number of ways. Starting with the DE fixed point equation

$$p\lambda\big(1-\rho(1-x)\big) = x \tag{16}$$

where $p$ designates the erasure probability of the BEC, we see that picking either the d.d. $\lambda$ or $\rho$ determines the other d.d. exactly. In this section, we make this notion precise and use it to expose some of the symmetries of c.a. LDPC codes.

A few definitions are needed to discuss things properly. Following the notation in [3], let $\mathcal{P}$ be the set of d.d. functions (i.e., functions $f$ with non-negative power series expansions around zero which satisfy $f(0) = 0$ and $f(1) = 1$); this set is defined by

$$\mathcal{P} \triangleq \left\{ f : f(x) = \sum_{k=1}^{\infty} f_k x^k,\ x \in [0,1], \quad f_k \geq 0,\ f(0) = 0,\ f(1) = 1 \right\}.$$

Let $\mathcal{T}$ be an operator which transforms invertible functions $f : [0,1] \to [0,1]$ according to the rule

$$\mathcal{T}f(x) \triangleq 1 - f^{-1}(1-x)$$

where $f^{-1}$ is the inverse function of $f$. The function $\mathcal{T}f$ is well-defined on $[0,1]$ for any function $f$ which is strictly monotonic on this interval, and therefore for any function in $\mathcal{P}$. We will say that two d.d. functions $f$ and $g$ are *matched* if $\mathcal{T}f = g$ (since $\mathcal{T}^2 f \equiv f$, the equality $\mathcal{T}f = g$ implies that $\mathcal{T}g = f$). Finally, let $\mathcal{A}$ be the set of all functions $f \in \mathcal{P}$ such that $\mathcal{T}f \in \mathcal{P}$, i.e.,

$$\mathcal{A} \triangleq \Big\{ f : f \in \mathcal{P} \,,\ \mathcal{T}f \in \mathcal{P} \Big\}.$$

The connection with LDPC codes is that finding some $f \in \mathcal{A}$ is typically the first step towards proving that $(f, \mathcal{T}f)$ is a c.a. d.d. pair. Truncation and normalization issues which depend on the erasure probability of the BEC must also be considered. When $p = 1$, many of these issues disappear, so we denote the set of d.d. pairs which satisfy (16) by

$$\begin{aligned} \mathcal{C}_{\text{LDPC}} &\triangleq \Big\{ (\lambda,\rho) \in \mathcal{P} \times \mathcal{P} \mid \ \lambda\big(1-\rho(1-x)\big) = x \Big\} \\ &= \Big\{ (\lambda,\rho) \mid \ \lambda \in \mathcal{A},\ \rho = \mathcal{T}\lambda \Big\}. \end{aligned}$$

The *symmetry property* of c.a. LDPC codes (with rate 0) asserts that

$$(\lambda,\rho) \in \mathcal{C}_{\text{LDPC}} \xleftrightarrow{\text{symmetry}} (\rho,\lambda) \in \mathcal{C}_{\text{LDPC}}. \tag{17}$$

One can prove this result by transforming (16) when $p = 1$. First, we let $x = 1 - \rho^{-1}(1-y)$, which gives

$$\lambda(y) = 1 - \rho^{-1}(1-y).$$

Then we rewrite this expression as

$$\rho\big(1-\lambda(y)\big) = 1 - y$$

and let $y = 1 - z$ to get

$$\rho\big(1-\lambda(1-z)\big) = z.$$

Comparing this with the DE fixed point equation (16) when $p = 1$ shows the symmetry between $\lambda$ and $\rho$.

## 3.2 Symmetry Properties of ARA Codes

The decoding of an ARA code can be broken into two stages. The first stage transforms the ARA code into an equivalent LDPC code via graph reduction, and the second stage decodes the LDPC code. This allows us to describe the symmetry property of c.a. ARA codes in terms of the symmetry property of c.a. LDPC codes. First, we introduce notation which allows us to express compactly the effect of graph reduction on an arbitrary d.d. from the edge perspective (see (4), (11) and (12)). For $f \in \mathcal{P}$, let us define

$$\widetilde{f}_p(x) \triangleq \frac{(1-p)^2\, f(x)}{\left(1 - \frac{p\int_0^x f(t)dt}{\int_0^1 f(t)dt}\right)^2}. \tag{18}$$

This allows the graph reduction of an ARA code to be interpreted as a mapping $\mathcal{G}_{\text{ARA}}$ from an ARA d.d. pair to an LDPC d.d. pair which can be expressed as

$$(\lambda,\rho) \overset{\mathcal{G}_{\text{ARA}}}{\rightleftarrows} \left(\widetilde{\lambda}_{1-p}, \widetilde{\rho}_p\right).$$

The inverse of the graph reduction mapping is represented by a dashed arrow because this inverse mapping, while always well-defined, does not necessarily preserve the property of having a non-negative power series expansion around zero.

Referring to ensembles of ARA codes, the set of d.d. pairs which satisfy the DE fixed point equation (8) is given by

$$\mathcal{C}_{\text{ARA}}(p) \triangleq \left\{ \, (\lambda, \rho) \in \mathcal{P} \times \mathcal{P} \mid \; \widetilde{\lambda}_{1-p}\big(1 - \widetilde{\rho}_p(1-x)\big) = x \, \right\}$$

where the equivalence to (8) follows from (11), (12) and (18).

The symmetry between the bit and check degree distributions of c.a. ARA ensembles follows from the symmetry relationship in (17), and the equivalence between a d.d. pair $(\lambda, \rho)$ for ARA codes and the d.d. pair $(\widetilde{\lambda}_{1-p}, \widetilde{\rho}_p)$ for LDPC codes of zero-rate.

The complete symmetry relationship for c.a. ARA ensembles over the BEC is therefore given in the following diagram:

$$\begin{array}{ccc}
(\lambda, \rho) \in \mathcal{C}_{\text{ARA}}(p) & \xleftrightarrow{\text{ARA symmetry}} & (\rho, \lambda) \in \mathcal{C}_{\text{ARA}}(1-p) \\
\mathcal{G}_{\text{ARA}} \big\downarrow \, \vdots & & \vdots \, \big\downarrow \mathcal{G}_{\text{ARA}} \\
\Big(\widetilde{\lambda}_{1-p}, \widetilde{\rho}_p\Big) \in \mathcal{C}_{\text{LDPC}} & \xleftrightarrow{\text{LDPC symmetry}} & \Big(\widetilde{\rho}_p, \widetilde{\lambda}_{1-p}\Big) \in \mathcal{C}_{\text{LDPC}}
\end{array}$$

The inverse of the graph reduction mapping is represented by the dashed arrow because this inverse transformation is only valid if it is known ahead of time that the power series expansions of $\lambda$ and $\rho$ are non-negative. It turns out that this symmetry is very useful in order to generate new d.d. pairs which satisfy the DE equality in (13). An alternative way to show this symmetry explicitly is rewriting (13)

$$\widetilde{\lambda}_{1-p}\big(1 - \widetilde{\rho}_p(x)\big) = x$$

and using the symmetry property (17) for LDPC codes to rewrite it as

$$\widetilde{\rho}_p\big(1 - \widetilde{\lambda}_{1-p}(x)\big) = 1 - x.$$

From (11) and (12), the expansion of the last equation gives

$$\frac{(1-p)^2 \, \rho\left(1 - \frac{p^2 \, \lambda(x)}{\big(1-(1-p)L(x)\big)^2}\right)}{\left(1 - p \, R\left(1 - \frac{p^2 \, \lambda(x)}{\big(1-(1-p)L(x)\big)^2}\right)\right)^2} = 1 - x. \tag{19}$$

Since the swapping $L(x) \leftrightarrow R(x)$, $\lambda(x) \leftrightarrow \rho(x)$, $p \leftrightarrow 1-p$, and $x \leftrightarrow 1-x$ maps this equation back to (8), then we can take any d.d. pair $(\lambda, \rho)$ which satisfies (8) for $p = p^*$ and swap $\lambda$ with $\rho$ (and hence, $L$ and $R$ are also swapped) to get a new d.d. pair which satisfies (19) for $p = 1 - p^*$ (equations (8) and (19) should be satisfied for all $x \in [0, 1]$, so switching between $x$ and $1 - x$ has no relevance).

### 3.3 Symmetry Properties of NSIRA Codes

Now, we consider the graph reduction process and symmetry properties of non-systematic irregular repeat-accumulate (NSIRA) codes (for preliminary material on NSIRA codes, the reader is referred to [8, Section 2]). In this respect, we introduce a new ensemble of codes which we call "Accumulate-LDPC" (ALDPC) codes. These codes are the natural image of NSIRA codes under the symmetry transformation. In fact, this ensemble was discovered by applying the symmetry transformation to previously known c.a. code ensembles. Their decoding graph can be constructed from the ARA decoding graph (see Fig. 2) by removing the bottom accumulate structure.

Since an NSIRA code has no accumulate structure attached to the "punctured bit" nodes, the graph reduction process affects only the d.d. of the "parity-check 2" nodes. Therefore, graph reduction acts as a mapping $\mathcal{G}_{\text{NSIRA}}$ from the NSIRA d.d. pair $(\lambda, \rho)$ to the LDPC d.d. pair $(\lambda, \widetilde{\rho}_p)$. This yields that for ensembles of NSIRA codes, the set of d.d. pairs which satisfy the DE fixed point equation is given by

$$\mathcal{C}_{\text{NSIRA}}(p) \triangleq \Big\{ (\lambda, \rho) \in \mathcal{P} \times \mathcal{P} \mid \; \lambda\big(1 - \widetilde{\rho}_p(1-x)\big) = x \Big\}.$$

An ALDPC code has no accumulate structure attached to the "parity-check 2" nodes, and therefore the graph reduction process only affects the d.d. of the "punctured bit" nodes. Hence, graph reduction acts as a mapping $\mathcal{G}_{\text{ALDPC}}$ from the ALDPC d.d. pair $(\lambda, \rho)$ to the LDPC d.d. pair $(\widetilde{\lambda}_{1-p}, \rho)$. For ALDPC ensembles, the set of d.d. pairs which satisfy the DE fixed point equation is therefore given by

$$\mathcal{C}_{\text{ALDPC}}(p) \triangleq \Big\{ (\lambda, \rho) \in \mathcal{P} \times \mathcal{P} \mid \widetilde{\lambda}_{1-p}\big(1 - \rho(1-x)\big) = x \Big\}.$$

The symmetry between c.a. ensembles of NSIRA and ALDPC codes over the BEC follows from the symmetry relationship in (17), the equivalence between a d.d. pair $(\lambda, \rho)$ for NSIRA codes and the d.d. pair $(\lambda, \widetilde{\rho}_p)$ for LDPC codes, and the equivalence between a d.d. pair $(\lambda, \rho)$ for ALDPC codes and the d.d. pair $(\widetilde{\lambda}_{1-p}, \rho)$ for LDPC codes.

The symmetry relationship between c.a. NSIRA and ALDPC ensembles over the BEC is therefore given in the following diagram:

$$\begin{array}{ccc}
(\lambda, \rho) \in \mathcal{C}_{\text{NSIRA}}(p) & \xleftrightarrow{\text{symmetry}} & (\rho, \lambda) \in \mathcal{C}_{\text{ALDPC}}(1-p) \\
\mathcal{G}_{\text{NSIRA}} \big\downarrow \big\uparrow & & \big\uparrow \big\downarrow \mathcal{G}_{\text{ALDPC}} \\
(\lambda, \widetilde{\rho}_p) \in \mathcal{C}_{\text{LDPC}} & \xleftrightarrow{\text{LDPC symmetry}} & (\widetilde{\rho}_p, \lambda) \in \mathcal{C}_{\text{LDPC}}
\end{array}$$

As before, the inverse of each graph reduction mapping is represented by a dashed arrow because this inverse transformation is only valid if it is known ahead of time that the power series expansions of $\lambda$ and $\rho$ are non-negative.

### 3.4 Connections with Forney's Transform

In [17], Forney introduces a graph transformation which maps the factor graph of any group code to the factor graph of the dual group code. For factor graphs of binary linear codes which only have equality and parity constraints (i.e., no trellis constraints), this operation is equivalent to swapping equality and parity constraints (e.g., bit nodes and check nodes). Forney's approach represents observations by half-edges, and these remain attached to the original node even though the nature of that node has changed. For example, Forney's transform maps an LDPC code with parity-check matrix $H$ to a low-density generator-matrix (LDGM) code with generator matrix $H$ and the half-edges attached to the bit nodes of the LDPC code are attached to the parity-check nodes of the LDGM code.

Using Forney's transform, we see that the swapping of $\lambda$ and $\rho$ described by our symmetry mappings actually transforms the original code ensemble into the dual code ensemble. Let the design rate of the original ensemble be $R$, then the design rate of the dual ensemble is $1-R$. This means that if we want to have any chance of achieving capacity, we must also map the channel erasure probability $p$ to $1-p$. Therefore, our symmetry relationships show that ARA, NSIRA, and ALDPC ensembles which are c.a. on BEC under iterative decoding also have dual ensembles which are c.a. on the BEC under iterative decoding.[1] In light of the Area Theorem and its relationship to the dual code [18], this result is not entirely surprising. Still, we had not considered the possibility that c.a. ensembles might automatically define c.a. dual ensembles.

Finally, we note that the basic structure of ARA codes is preserved under Forney's transform. In particular, this means that we can construct self-dual ARA codes (i.e., rate $\frac{1}{2}$) by choosing the square connection matrix between the "punctured bit" nodes and the "parity-check 2" nodes to be symmetric.

# 4 Capacity-Achieving ARA Ensembles for the BEC

This section serves as a preparatory step towards the construction of explicit c.a. ARA ensembles for the BEC, whose decoding complexities stay bounded as the gap to capacity vanishes. Later in Section 5 we will present explicit constructions of bit-regular and check-regular ARA ensembles which are based on a similar approach due to the symmetry properties provided in the previous section. Section 6 introduces another approach for the construction of c.a. ensembles of ARA codes with bounded complexity over the BEC. The concepts used for these constructions are based on the symmetry properties in the previous section, and the material presented in this section.

## 4.1 A Starting Point for Constructing Capacity-Achieving ARA Ensembles

Using the tilted degree distributions after graph reduction given in Section 2.3.2, we apply the DE equation for LDPC codes (13) to derive c.a. sequences. This property is proved in the following lemma.

**Lemma 1.** If equality (13) is satisfied for all $x \in [0,1]$, then the design rate of the corresponding ensemble of ARA codes is equal to the capacity of the BEC.

*Proof.* From the condition in (13), it follows that

$$1 - \widetilde{\lambda}^{-1}(1-x) = \widetilde{\rho}(x). \tag{20}$$

Since $\widetilde{\lambda}(0) = 0$, $\widetilde{\lambda}(1) = 1$ and $\widetilde{\lambda}$ is a monotonic differentiable function on the interval $[0,1]$, then

[1]To be precise, we actually need to consider sequences of ensembles which are c.a. and relate them to sequences of dual ensembles. This distinction is rather cumbersome and does not cause problems in this case.

from (20), the substitution $x = \lambda(u)$ and integration by parts give

$$\begin{aligned}
\int_0^1 \widetilde{\rho}(x)\,dx &= \int_0^1 \left(1 - \widetilde{\lambda}^{-1}(1-x)\right) dx \\
&= 1 - \int_0^1 \widetilde{\lambda}^{-1}(x)\,dx \\
&= 1 - \int_0^1 u\widetilde{\lambda}'(u)\,du \\
&= 1 - \left[u\,\widetilde{\lambda}(u)\Big|_0^1 - \int_0^1 \widetilde{\lambda}(u)du\right] \\
&= \int_0^1 \widetilde{\lambda}(u)du. \qquad (21)
\end{aligned}$$

From (3), (11), and the equalities $L(0) = 0$ and $L(1) = 1$, we get

$$\begin{aligned}
\int_0^1 \widetilde{\lambda}(x)\,dx &= \int_0^1 \frac{p^2\lambda(x)}{\left(1-(1-p)L(x)\right)^2}\,dx \\
&= \frac{1}{L'(1)} \int_0^1 \frac{p^2 L'(x)}{\left(1-(1-p)L(x)\right)^2}\,dx \\
&= \frac{1}{L'(1)} \int_0^1 \frac{p^2 du}{\left(1-(1-p)u\right)^2} \\
&= \frac{p}{L'(1)}. \qquad (22)
\end{aligned}$$

Similarly, from (3), (12) and the equalities $R(0) = 0$ and $R(1) = 1$, we obtain

$$\int_0^1 \widetilde{\rho}(x)\,dx = \frac{1-p}{R'(1)}. \qquad (23)$$

By combining (21)–(23), we obtain the equality

$$\frac{L'(1)}{R'(1)} = \frac{p}{1-p} \qquad (24)$$

and hence, from (5), the design rate of the ensemble of ARA codes is equal to $1-p$ (i.e., the ensemble achieves the capacity of the BEC). □

Now, consider the DE fixed point equation (13) (or equivalently (8)). Using this equation, we see that the condition $\rho(0) = 0$ (i.e., no degree-1 "parity-check 2" nodes) is necessary and sufficient to guarantee that (13) is always satisfied at $x = 1$. Likewise, the condition $\lambda(0) = 0$ (i.e., no degree-1 "punctured-bit" nodes) is necessary and sufficient to guarantee that (13) is always satisfied at $x = 0$. From Lemma 1, we conclude that if there exists a d.d. pair $(\lambda, \rho)$ with bounded average degree that satisfies (8), then there is a c.a. sequence of ARA ensembles with bounded complexity for the BEC. This conclusion is also based on the truncation discussed in the next section.

## 4.2 Truncating Degree Distributions

After finding a d.d. pair which satisfies the DE equation with equality, a suitable truncation can be used to exhibit a sequence of ensembles that achieves capacity. Consider, for example, a sequence of d.d. pairs $\{(\lambda^{(M)}, \rho^{(M)})\}_{M\in\mathbb{N}}$ indexed by the maximum degree $M$. Since the effect of each truncation is negligible as $M$ goes to infinity, Lemma 1 shows that the design rate approaches capacity in this case. If the truncations are chosen properly, then we can also show that each truncated d.d. pair in the sequence has no DE fixed points for $x \in (0, 1]$. This implies that, for any $\delta > 0$, there exists a block length $n_0(\delta, M)$ such that the probability of decoding failure is less than $\delta$ for all block lengths $n > n_0$.

Since our DE equations depend on both the edge and node degree distributions, the truncation must be chosen carefully to simultaneously bound both. For the check d.d., we want modified degree distributions $\hat{R}$ and $\hat{\rho}$ such that $\hat{R}(x) > R(x)$ and $\hat{\rho}(x) > \rho(x)$ for $x \in [0, 1)$. In particular, we replace large degree checks by degree-1 checks and this gives

$$\begin{aligned}\hat{\rho}(x) &= \left(\rho_1 + \sum_{i=M+1}^{\infty} \rho_i\right) + \sum_{i=2}^{M} \rho_i x^{i-1} \\ \hat{R}(x) &= \frac{\int_0^x \hat{\rho}(t)dt}{\int_0^1 \hat{\rho}(t)dt}.\end{aligned}$$

This truncation was introduced in [8] and proven to satisfy the desired conditions.

For the bit d.d., we want truncated degree distributions $\hat{L}$ and $\hat{\lambda}$ that satisfy $\hat{L}(x) < L(x)$ and $\hat{\lambda}(x) < \lambda(x)$ for $x \in (0, 1]$. In this case, we replace large degree bits by pilot bits (e.g., these bits are forced to zero and known at receiver); this gives

$$\begin{aligned}\hat{\lambda}(x) &= \sum_{i=1}^{M} \lambda_i x^{i-1} \\ \hat{L}(x) &= \sum_{i=1}^{M} L_i x^i.\end{aligned}$$

We note that this truncation satisfies the desired conditions (as long as $L_i > 0$ for some $i > M$) because it simply removes positive terms.

### 4.3 Encoding and Decoding Complexity

When transmission takes place over a BEC, the encoding/decoding complexity under iterative message-passing decoding is defined to be the average number of edges per information bit in the Tanner graph of the code (see Fig. 2 in p. 4). The motivation for measuring the complexity in this way is because the encoder and the iterative decoder can be both designed to use every edge in the graph exactly one time (due to the absolute reliability of information provided by the BEC).

From the Tanner graph of ARA codes in Fig. 2, it can be verified that the encoding complexity ($\chi_{\text{E}}$) and the decoding complexity ($\chi_{\text{D}}$) are both equal to

$$\chi_{\text{E}} = \chi_{\text{D}} = 3 + L'(1) + \frac{2(1-R)}{R} \tag{25}$$

where $R$ is the design rate of the ensemble.

The complexity of NSIRA codes can also be computed from Fig. 2 by ignoring the accumulate structure for the systematic bits. This shows that

$$\chi_{\text{E}} = \chi_{\text{D}} = L'(1) + \frac{2}{R}. \tag{26}$$

Likewise, the complexity of ALDPC codes can be computed from Fig. 2 by ignoring the accumulate structure for the parity bits. This shows that

$$\chi_{\mathrm{D}} = \frac{3 + L'(1)}{R}. \tag{27}$$

In general, the encoding complexity of ALDPC codes does not grow linearly with the block length because it requires the encoding of an LDPC code which can be quadratic in the block length. We can, however, apply fast encoding methods for c.a. LDPC codes [16] to c.a. ALDPC codes. These methods will result in linear-time encoding algorithms for c.a. ALDPC codes.

### 4.4 The Effect of Puncturing

Puncturing is a well-known technique that allows one to design for one code rate and adaptively increase that rate to match channel conditions. Strictly speaking, we note that punctured ARA ensembles are no longer systematic because some information bits may not be transmitted as a result of the puncturing. This technique, however, can be used to extend the range of $p$ for which certain d.d. pairs are c.a. with bounded complexity.

For example, consider any code construction which is provably c.a. with bounded complexity for $p > p_0$ (e.g., the check-regular ARA ensemble which will be introduced in Section 5). This construction can be made to achieve capacity for $0 < p < 1$ simply by puncturing bits at random before transmission (i.e., all bits have the same puncturing rate). Let $\alpha$ be the fraction of bits transmitted, then the effective erasure rate of the channel is given by $p_{\text{eff}} = 1 - \alpha(1-p)$. Picking $\alpha < 1 - p_0$ guarantees that $p_{\text{eff}} > p_0$ and that the ensemble achieves capacity. This operation does increase the complexity by a factor of $\frac{1}{\alpha}$ because the punctured bits must be retained as part of the decoding graph. We apply this method in some computer simulations to increase the code rate of a particularly good ensemble of rate $\frac{1}{2}$ codes.

Codes with two classes of bits (e.g., ARA codes) may also benefit from asymmetric puncturing of the two classes. For example, puncturing all of the systematic bits of an ARA code converts that code into a NSIRA code [8]. So we find that sending a fraction $\alpha$ of the systematic bits of an ARA code gives a smooth transition between ARA codes and NSIRA codes for $\alpha \in [0, 1]$.

## 5 Bit-Regular and Check-Regular Capacity-Achieving Ensembles with Bounded Complexity for the BEC

This section gives explicit constructions of c.a. ARA ensembles for the BEC, which are either bit-regular or check-regular. As will be observed, these ensembles possess bounded complexity (per information bit) as the gap to capacity vanishes.

The symmetry property in Section 3.2 allows one for example to design an ensemble of high rate ARA codes, and get automatically (by switching between the pair of degree distributions) a new ensemble of ARA codes which is suited for low rate applications. We will rely on this symmetry property in Section 5.2 when we transform a bit-regular ARA ensemble designed for a BEC with erasure probability $p \in (0, p^*]$ into a check-regular ensemble designed for $p \in [1 - p^*, 1)$. We also rely on the fact that the method in Section 5.1 for computing the function $R$ given the function $L$ can be easily inverted using the symmetry property. This means that given an algorithm to solve for $R(x)$ in terms of $L(x)$ for a certain $p_0$, the inverse algorithm which solves $L(x)$ in terms of $R(x)$ is exactly the same, except that $p_0$ is replaced by $1 - p_0$.

## 5.1 Solving for $R(x)$ in terms of $L(x)$

Given $L(x)$, we start with the calculation of $\lambda(x) = \frac{L'(x)}{L'(1)}$. Then $\widetilde{\lambda}(x)$ is calculated from (11), and $\widetilde{\rho}(x) = 1 - \widetilde{\lambda}^{-1}(1-x)$ is calculated from (13). Combining (3) and (12) gives

$$\widetilde{\rho}(x) = \frac{(1-p)^2}{R'(1)} \frac{R'(x)}{\big(1 - pR(x)\big)^2}$$

and by integrating both sides of this equation, we get

$$\int_0^x \widetilde{\rho}(t)\, dt = \frac{(1-p)^2}{R'(1)} \frac{R(x)}{1-pR(x)}. \tag{28}$$

Since $R(1) = 1$, substituting $x = 1$ in the last equality and solving for $R'(1)$ gives

$$R'(1) = \frac{1-p}{\displaystyle\int_0^1 \widetilde{\rho}(t)\, dt}.$$

By substituting the last equality in (28), we get

$$Q(x) = \frac{(1-p)R(x)}{1-pR(x)} \tag{29}$$

where

$$Q(x) \triangleq \frac{\int_0^x \widetilde{\rho}(t)\, dt}{\int_0^1 \widetilde{\rho}(t)\, dt}. \tag{30}$$

Using the fact that $y = \frac{z}{1-pz} \implies z = \frac{y}{1+py}$, we solve (29) for $R(x)$ and get

$$R(x) = \frac{Q(x)}{1-p+pQ(x)}. \tag{31}$$

Combining (3), (30) and (31) gives

$$\begin{aligned} \rho(x) &= \frac{R'(x)}{R'(1)} \\ &= \frac{\widetilde{\rho}(x)}{\big(1-p+pQ(x)\big)^2}. \end{aligned} \tag{32}$$

As long as we have $\widetilde{\rho}(1) = 1$, then evaluating (32) at $x = 1$ gives $\rho(1) = 1$. Therefore, there is no need to truncate the power series of $\rho$. As we noted above, a very similar approach can be applied to solve for $L(x)$ in terms of $R(x)$; due to the symmetry property, one can simply apply the above procedure to a parity-check d.d. $R(x)$ with an erasure probability of $1-p$.

## 5.2 Bit-Regular and Check-Regular Capacity-Achieving ARA Ensembles

The symmetry between bit-regular and check-regular c.a. ensembles of ARA codes follows from the symmetry properties presented in Section 3.2, so we choose to focus on a bit-regular ARA ensemble. Let $\lambda(x) = x^2$, so $L(x) = x^3$, and from (11)

$$\widetilde{\lambda}(x) = \frac{p^2 x^2}{\big(1-(1-p)x^3\big)^2}.$$

Based on (13), we get

$$\begin{aligned}\widetilde{\rho}^{-1}(x) &= 1-\widetilde{\lambda}(1-x) \\ &= 1-\frac{p^2(1-x)^2}{\big(1-(1-p)(1-x)^3\big)^2}.\end{aligned} \tag{33}$$

This is exactly [8, Eq. (39)] with $p$ replaced by $1-p$ and the $\rho$ switched with $\lambda$. Therefore, we obtain from [8, Theorem 2] that the tilted d.d. $\widetilde{\rho}$ has the form

$$\widetilde{\rho}(x) = 1 + \frac{2(1-p)(1-x)^2 \sin\left(\frac{1}{3}\arcsin\left(\sqrt{-\frac{27(1-p)(1-x)^{\frac{3}{2}}}{4p^3}}\right)\right)}{\sqrt{3}\,p^4\left(-\frac{(1-p)(1-x)^{\frac{3}{2}}}{p^3}\right)^{\frac{3}{2}}}. \tag{34}$$

Following the procedure of Section 5.1, starting from (30), gives (after some calculus)

$$Q(x) = \frac{3(x-1)\widetilde{\rho}(x)}{p} + \frac{1-\big(1-\widetilde{\rho}(x)\big)^3}{1-(1-p)\big(1-\widetilde{\rho}(x)\big)^3} \tag{35}$$

where the calculations leading to the expression for $Q$ are detailed in Appendix 8. Substituting $Q$ in (35) into (31) gives the following expression of the d.d. $R$:

$$\begin{aligned}R(x) &= \frac{1}{p}\left(1-\frac{1-p}{1-p+pQ(x)}\right) \\ &= \frac{1}{p}\left(1-\frac{1}{1+\frac{3(x-1)\widetilde{\rho}(x)}{1-p}+\frac{p}{1-p}\,\frac{1-\left(1-\widetilde{\rho}(x)\right)^3}{1-(1-p)\left(1-\widetilde{\rho}(x)\right)^3}}\right)\end{aligned} \tag{36}$$

where the function $\widetilde{\rho}$ is given in (34). It was verified numerically that for $p \leq 0.384$, the first 300 coefficients of the power series expansion of the d.d. $R$ are non-negative; in Appendix C.3.2, we prove that if $p \leq 0.26$, then $R$ has indeed a power series expansion about $x = 0$ whose all coefficients are non-negative. According to Lemma 1 (see p. 13), it also holds in general that any d.d. pair satisfying (8) has a design rate equal to the capacity of the BEC. It therefore appears that the d.d. pair above characterizes a c.a. ensemble of bit-regular ARA codes over the BEC; the capacity of the BEC is achieved with bounded complexity for rates greater than 0.616. We note that the convergence speed of the degree distribution for the parity-check nodes is relatively fast. As an example, for $p = 0.3$, the fraction of check nodes with degree less than 32 is equal to 0.968.

Using the symmetry between $\widetilde{\lambda}$ and $\widetilde{\rho}$ (see Section 3), this also implies that for rates less than 0.384, the ensemble of check-regular ARA codes with $R(x) = x^3$ achieves capacity over the BEC with bounded complexity. Based on the symmetry property for c.a. ensembles of ARA codes, the d.d. $L$ for the check-regular ARA ensemble is obtained from the d.d. $R$ for the bit regular ARA ensembles when $p$ is replaced by $1-p$. From (36) and the symmetry property, the d.d. $L$ for the check-regular ARA ensemble which corresponds to $R(x) = x^3$ has the form

$$L(x) = \frac{1}{1-p}\left(1-\frac{1}{1+\frac{3(x-1)\widetilde{\lambda}(x)}{p}+\frac{1-p}{p}\,\frac{1-\left(1-\widetilde{\lambda}(x)\right)^3}{1-p\left(1-\widetilde{\lambda}(x)\right)^3}}\right) \tag{37}$$

where for this ensemble

$$\widetilde{\lambda}(x) = 1 + \frac{2p(1-x)^2 \sin\left(\frac{1}{3}\arcsin\left(\sqrt{-\frac{27p(1-x)^{\frac{3}{2}}}{4(1-p)^3}}\right)\right)}{\sqrt{3}\,(1-p)^4\left(-\frac{p(1-x)^{\frac{3}{2}}}{(1-p)^3}\right)^{\frac{3}{2}}}. \tag{38}$$

Note that the d.d. $\widetilde{\lambda}$ in (38) is obtained by replacing $p$ by $1-p$ in the RHS of (34), which finally gives the d.d. function introduced in [8, Eq. (15)].

### 5.3 Capacity-Achieving ALDPC Ensembles

Using the symmetry relationship between NSIRA and ALDPC ensembles from Section 3.3, we find that we already have from [8, Theorems 1 and 2] two c.a. ensembles of ALDPC codes. These ensembles are based on the bit-regular and check-regular NSIRA ensembles of [8]. This was also observed independently by Hsu and Anastasopoulos [12].

Using symmetry, the check-regular NSIRA ensemble gives a bit-regular ALDPC ensemble which provably achieves capacity with bounded complexity for $p \in (0,1)$. Since d.d. for small $p$ has long tails, one can also use random puncturing to increase the effective erasure rate of the channel, and therefore simplify code design. Similarly, the bit-regular NSIRA ensemble gives a check-regular ALDPC ensemble which provably achieves capacity with bounded complexity for $p \in \left[\frac{12}{13}, 1\right)$. In this case, random puncturing can be used to extend the valid range to $(0,1)$.

The replacement of $p$ by $1-p$ in [8, Eq. (15)] which corresponds to the right d.d. of the check-regular NSIRA ensemble, gives the following check d.d. for the bit-regular ALDPC ensemble with left degree of 3

$$\rho(x) = 1 + \frac{2(1-p)(1-x)^2 \sin\left(\frac{1}{3}\arcsin\left(\sqrt{-\frac{27(1-p)(1-x)^{\frac{3}{2}}}{4p^3}}\right)\right)}{\sqrt{3}\,p^4\left(-\frac{(1-p)(1-x)^{\frac{3}{2}}}{p^3}\right)^{\frac{3}{2}}}. \tag{39}$$

Likewise, the replacement of $p$ by $1-p$ in [8, Eq. (10)] which corresponds to the left d.d. of the bit-regular NSIRA ensemble, and the substitution $q=3$ gives the following bit d.d. for the check-regular ALDPC ensemble with right degree of 3

$$\lambda(x) = \frac{1-(1-x)^{\frac{1}{2}}}{\left[1-(1-p)\left(1-3x+2\left[1-(1-x)^{\frac{3}{2}}\right]\right)\right]^2}. \tag{40}$$

It is worth noting that the bit-regular ALDPC ensemble has minimum bit degree of 3. Therefore, truncating the check d.d. to finite maximum check degree makes the ensemble unconditionally stable. It also shows that minimum distance of should grow linearly with the block length because the minimum distance of LDPC codes with a minimum bit dgeree of 3 and a fixed maximum check degree grows linearly with the block length [19]. To prove this rigorously, however, one must also consider the effect of the accumulate structure on the minimum distance.

## 6 Capacity-Achieving Ensembles with Bounded Complexity for the BEC: Constructions Based on LDPC Codes

In this section, we introduce another way of constructing c.a. ensembles of ARA codes for the BEC. Rather then solving for the function $R$ in terms of the function $L$ (as in Section 5.1) or doing the inverse via the symmetry property, we consider here another natural way of searching for c.a. degree distributions. We start by choosing a candidate d.d. pair $(\widetilde{\lambda}, \widetilde{\rho})$ which satisfies equation (13) and test to see if it can be used to construct an ensemble of c.a. ARA codes. The testing process starts by mapping the tilted pair $(\widetilde{\lambda}, \widetilde{\rho})$ back to $(\lambda, \rho)$ via (11) and (12), and then testing the non-negativity of the resulting power series expansions of $\lambda$ and $\rho$.

Following the notation in Section 3.1, it enables one to rewrite (13) as $\widetilde{\rho} = \mathcal{T}\widetilde{\lambda}$ (so the tilted degree distributions $\widetilde{\lambda}$ and $\widetilde{\rho}$ are matched), and gives a compact description of capacity-achieving d.d. pairs of LDPC codes. We note that since $\mathcal{T}^2 f = f$ for an arbitrary function $f$ which has an inverse, then $f \in \mathcal{A}$ if and only if $\mathcal{T}f \in \mathcal{A}$. Based on (13), we obtain that we need to choose the tilted d.d. so that $\widetilde{\lambda} \in \mathcal{P}$ and also $\mathcal{T}\widetilde{\lambda} \in \mathcal{P}$, i.e., we need that the d.d. $\widetilde{\lambda}$ (or $\widetilde{\rho}$) both belong to the set $\mathcal{A}$. The reader is referred to [3, Lemma 1] which considers basic properties of the set $\mathcal{A}$ and the transformation $\mathcal{T}$.

So far, by choosing $\widetilde{\lambda} \in \mathcal{A}$ (or $\widetilde{\rho} \in \mathcal{A}$), we only know that both tilted d.d. have non-negative power series expansions. This property does not ensure that both of the original (i.e., non-tilted) d.d. $\lambda$ and $\rho$ also have non-negative power series expansions. Calculation of $\lambda$ and $\rho$ from the tilted d.d. $\widetilde{\lambda}$ and $\widetilde{\rho}$ is not straightforward since both equations involve the d.d. $L$ and $R$ which are the normalized integrals of the unknown $\lambda$ and $\rho$. In order to overcome this difficulty in solving the two integral equations, we suggest calculating the tilted d.d. pair w.r.t. the nodes of the graph using

$$\widetilde{L}(x) = \frac{\int_0^x \widetilde{\lambda}(t)\, dt}{\int_0^1 \widetilde{\lambda}(t)\, dt}\ , \quad \widetilde{R}(x) = \frac{\int_0^x \widetilde{\rho}(t)\, dt}{\int_0^1 \widetilde{\rho}(t)\, dt}. \tag{41}$$

The original d.d. pair w.r.t. the nodes (i.e., the original d.d. pair before the graph reduction) can be calculated from Eqs. (9) and (10). We obtain that

$$L(x) = \frac{\widetilde{L}(x)}{p + (1-p)\widetilde{L}(x)}\ , \quad R(x) = \frac{\widetilde{R}(x)}{1 - p + p\widetilde{R}(x)} \tag{42}$$

and then use equation (3) to find $(\lambda, \rho)$. The critical issue here is to verify whether the functions $L$ and $R$ have non-negative power series expansions.

### 6.1 Capacity-Achieving ARA Ensembles from Self-Matched LDPC Codes

It is easy to verify that the function

$$f(x) = \frac{(1-b)x}{1-bx}\ , \quad 0 < b < 1 \tag{43}$$

belongs to the set $\mathcal{A}$ and also $\mathcal{T}f = f$; in the case where $\mathcal{T}f = f$, the function $f$ is said to be self-matched. Therefore, based on (13), we examine here whether the choice $\widetilde{\lambda}(x) = \widetilde{\rho}(x) = \frac{(1-b)x}{1-bx}$ can be transformed into an ensemble of ARA codes whose degree distributions have non-negative power series expansions. From (41) and (42), we get

$$\widetilde{L}(x) = \widetilde{R}(x) = \frac{bx + \ln(1-bx)}{b + \ln(1-b)} \tag{44}$$

and from (42), we obtain that

$$L(x) = \frac{bx + \ln(1-bx)}{p\,[b + \ln(1-b)] + (1-p)\,[bx + \ln(1-bx)]} \tag{45}$$

$$R(x) = \frac{bx + \ln(1-bx)}{(1-p)\,[b + \ln(1-b)] + p\,[bx + \ln(1-bx)]}. \tag{46}$$

Since we started with the function $f$ in (43) which is self-matched, the resulting functions $L$ and $R$ in this approach are exactly the same, except that $p$ and $1-p$ are switched. In Appendix C.3, it is proved that the degree distributions $L$ and $R$ in (45) and (46), respectively, have non-negative power series expansion if and only if

$$\frac{1}{1 - \frac{13-\sqrt{61}}{9} \cdot \Big(b + \ln(1-b)\Big)} \leq p \leq 1 - \frac{1}{1 - \frac{13-\sqrt{61}}{9} \cdot \Big(b + \ln(1-b)\Big)}. \tag{47}$$

Fortunately, there exists a region of $(b, p)$ where this condition is satisfied. For the specification of this region, we use the Lambert W-function $W(x)$ which is defined to be the $w$-solution of the equation $we^w = x$; this function is real for $x > -\frac{1}{e}$. In the following we introduce and prove the following theorem:

**Theorem 1 (Ensembles of Self-Matched ARA Codes).** The ensemble of self-matched ARA codes, defined by the pair of degree distributions $(L, R)$ in (45) and (46), achieves the capacity of the BEC for any erasure probability $p \in (0, 1)$. This result is achieved under iterative message-passing decoding with *bounded complexity.*

The tails of the d.d. (i.e., the partial sums $\sum_{i=k}^{\infty} L_i$ and $\sum_{i=k}^{\infty} R_i$) *decay exponentially* like $O(b^k)$ where the parameter $b$ is given in terms of the Lambert W-function as

$$b = W\left(-e^{-\frac{13+\sqrt{61}}{12}\frac{1+|1-2p|}{1-|1-2p|}-1}\right) + 1. \tag{48}$$

The complexity, per information bit, of encoding and decoding is given by

$$\chi_{\rm E} = \chi_{\rm D} = \frac{3-p}{1-p} - \frac{b^2 p}{(1-b)[b + \ln(1-b)]}\ . \tag{49}$$

*Proof.* Referring to the pair of degree distribution $L$ and $R$ in (45) and (46), respectively, we need to obtain the necessary and sufficient conditions which ensure that these two function have non-negative power series expansion about $x = 0$. For a given value of $b$ in these degree distributions, it is proved in Appendix C.3.3 that this property is satisfied if and only if the inequality in (47) holds.

The encoding and decoding complexities of c.a. ensembles of ARA codes for the BEC are discussed in Section 4.3. Since our ensemble is c.a., then $R = 1 - p$ where $p$ designates the erasure probability of the BEC, and from (42) and (44)

$$L'(1) = p\,\widetilde{L}'(1) = -\frac{b^2 p}{(1-b)[b + \ln(1-b)]}\ .$$

Combining (25) with the last equality provides the expression in (49) for the complexity, per information bit, of encoding and decoding.

For fixed $p \in (0, 1)$, the complexity in (49) forms a monotonic increasing function of $b$ (which becomes unbounded as $b \to 1^-$). In order to minimize the encoding/decoding complexity, we wish to find the smallest value of $b$ in the interval $(0,1)$ so that the power series expansions about zero of the degree distributions $L$ and $R$ are both non-negative. For a fixed value of $p$, it is equivalent to solving for the minimal value of $b \in (0,1)$ which satisfies the condition in (47). This gives the equation

$$\frac{1}{1 - \frac{13-\sqrt{61}}{9} \cdot \Big(b + \ln(1-b)\Big)} = \min(p, 1-p)$$

which can be rewritten as

$$-b - \ln(1-b) = \frac{13+\sqrt{61}}{12} \, \frac{1+|1-2p|}{1-|1-2p|} \tag{50}$$

by using the equality

$$\min(p, 1-p) = \frac{1}{2} - \Big|\frac{1}{2} - p\Big|.$$

For $a \in \mathbb{R}$, the solution of the equation $-b - \ln(1-b) = a$ is given by $b = W(-e^{-1-a}) + 1$. To verify this, one needs to write the equation in the form $(b-1)e^{b-1} = -e^{-a-1}$, and rely on the definition of the Lambert W-function. Hence, the solution of equation (50) is given by the expression for $b$ in (48). For $p \in (0,1)$, the expression for $b$ in (48) achieves its global minimum at $p = \frac{1}{2}$, and its value is

$$b^* = W(-e^{-\frac{25+\sqrt{61}}{12}}) + 1 \approx 0.9304.$$

Eq. (48) therefore implies that for $0 < p < 1$, the parameter $b$ ranges in the interval $[b^*, 1)$; it achieves the value $b = b^*$ at $p = \frac{1}{2}$, and tends to 1 when $p$ approaches zero or unity.

The asymptotic behavior of the two d.d. pairs w.r.t. the nodes and the edges is derived in Appendix B.3, and is given by

$$L_k, R_k = O\left(\frac{b^k}{k \ln^2(k)}\right), \qquad \lambda_k, \rho_k = O\left(\frac{b^k}{\ln^2(k)}\right) \tag{51}$$

so the tails of the d.d. pair $(L, R)$ decay exponentially with $k$. □

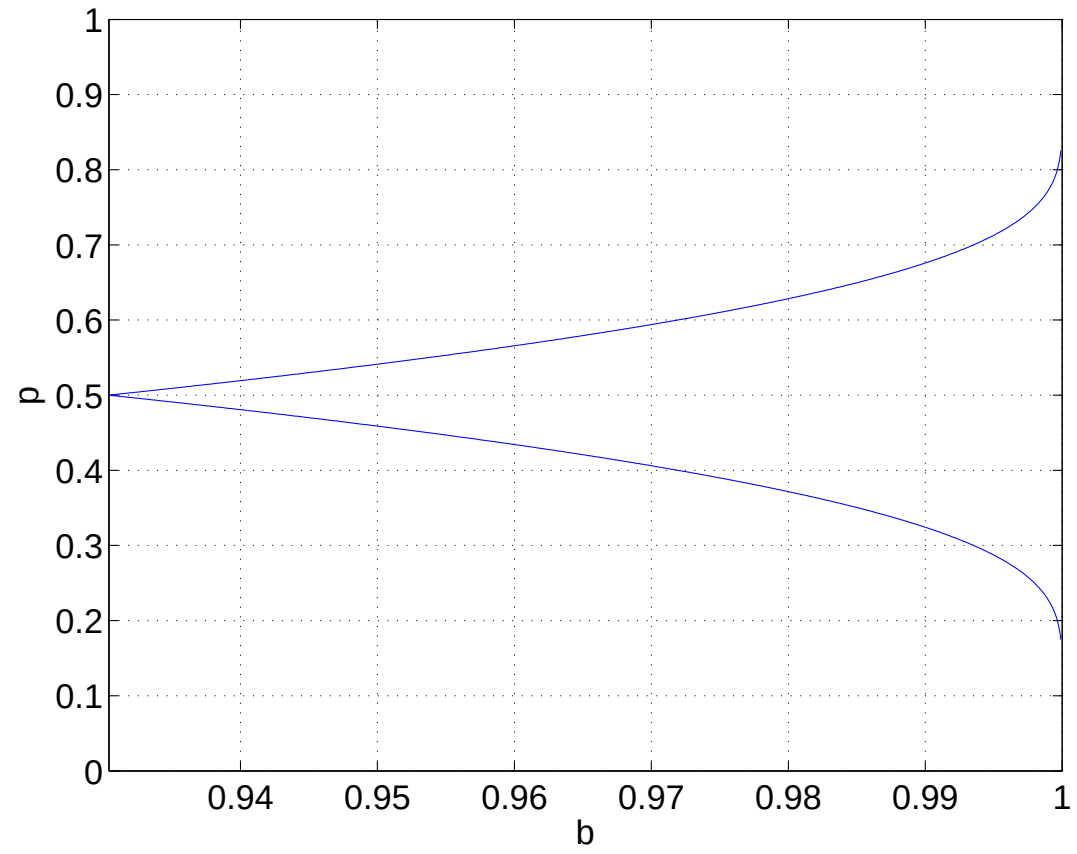

Figure 4: Valid values of $p$ as a function of the parameter $b$ in (45) and (46), so that the coefficients in the power series expansions around zero of $L(\cdot)$, $R(\cdot)$, $\lambda(\cdot)$ and $\rho(\cdot)$ are all non-negative. This region is determined by the inequalities in (47), and it is therefore bounded by the upper and lower curves in this figure.

The region $(b, p)$ characterized by the inequality in (47) is depicted in Fig. 4, and we point out that its width grows as $b$ gets closer to 1. Note that it follows from (47) that in the limit where $b \to 1^-$, the d.d. pair $L$ and $R$ have non-negative power series expansions for $0 < p < 1$. However, from (49), the complexity in the limit where $b \to 1^-$ becomes unbounded.

An efficient algorithm for the calculation of the d.d. pair in (45) and (46) w.r.t. the nodes of the graph, and the d.d. pair $(\lambda, \rho)$ w.r.t. the edges is given in Appendix B.2.

We believe the performance advantage of this ensemble over other c.a. ensembles is mainly due to the *exponential* decay of the d.d. coefficients, as given in (51).

### Numerical Results

In Fig. 5, we show the encoding and decoding complexity of the self-matched ARA ensemble introduced in this section (left plot), and the minimal value of $k$ such that the partial sums $\sum_{i=2}^{k} \lambda_i$ and $\sum_{i=2}^{k} \rho_i$ exceed 0.95 (right plot).

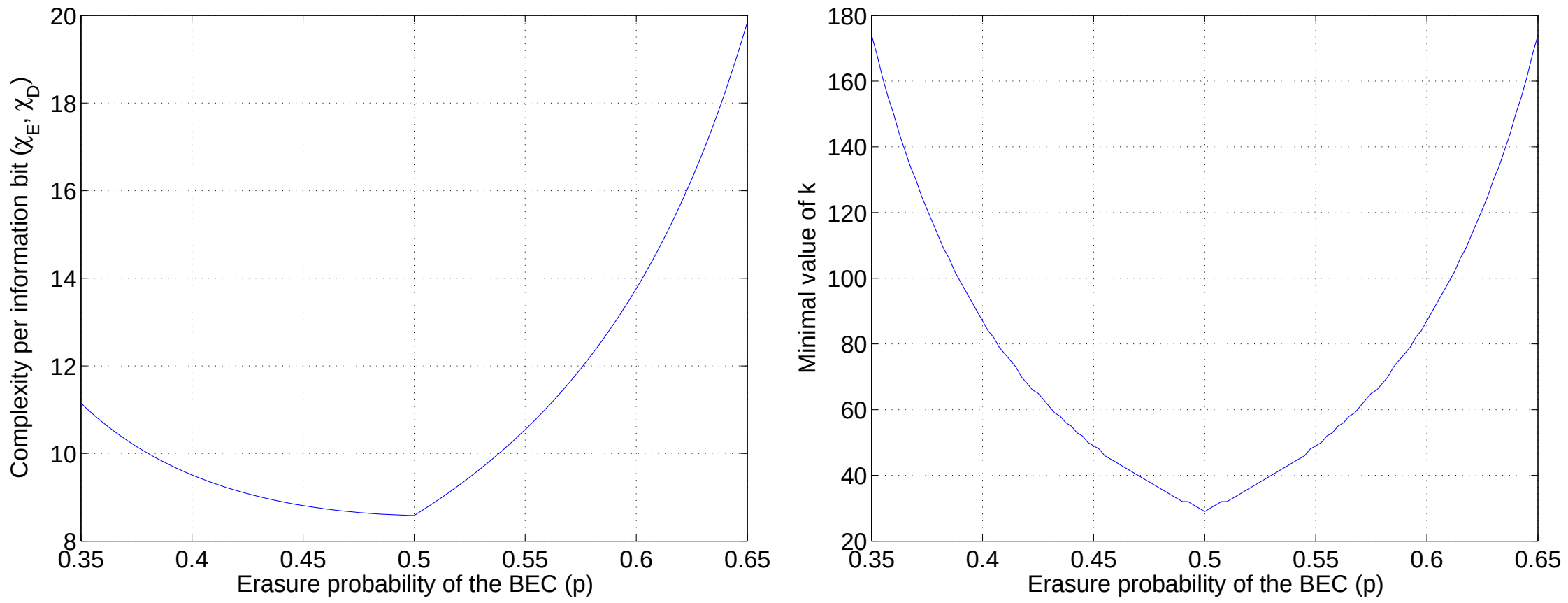

Figure 5: Left plot: Encoding/decoding complexity per information bit of the considered ensemble of ARA codes. The complexity is given in (49) as a function of the erasure probability ($p$) of the BEC, and to this end, we choose the parameter $b$ according to equation (48) for minimizing the complexity. Right plot: A plot of the minimal value of $k$ as a function of the erasure probability ($p$) so that the partial sums $\sum_{i=2}^{k} \lambda_i$ and $\sum_{i=2}^{k} \rho_i$ both exceed 0.95. We choose the parameter $b$ according to equation (48) which also minimizes $k$ for a fixed value of $p$.

Both of these sums converge to 1 as $k$ goes to infinity, and the convergence time is measured by the minimal value of $k$ where these partial sums exceed a threshold which is close to 1 (e.g., 0.95). We note that, like the complexity, the convergence time of these partial sums is increasing with $b$. Therefore, the choice we made according to equation (48) minimizes both quantities simultaneously. Notice that the complexity and convergence time are rather small for $0.35 \le p \le 0.65$. Both of these quantities achieve their minimal value at $p = \frac{1}{2}$ since the value of $b$ required by (48) is also minimized. We also note that the minimal value of $k$ is symmetric around $p = \frac{1}{2}$ (see lower plot) while the complexity is asymmetric (see upper plot). The reason for the symmetry property around $p = \frac{1}{2}$ in the lower plot of Fig. 5 is because the replacement of $p$ by $1 - p$ yields the same value of $b$ in (48); actually, the replacement of $p$ by $1 - p$ yields the same d.d. pair, except that $\{\lambda_i\}_{i \ge 2}$ and $\{\rho_i\}_{i \ge 2}$ are switched (this follows directly from (45) and (46)).

For $p = 0.50$ and $p = 0.60$, the convergence rates of the degree distributions w.r.t. the nodes and edges of the graph are shown in Fig. 6. The value of the parameter $b$ is determined by (48), and the corresponding complexity of encoding and decoding is equal to 8.585 and 13.776, respectively (the complexity is therefore relatively small in both cases). The results for a BEC with erasure probability $p = 0.50$ (i.e., a design rate of $\frac{1}{2}$) are particularly encouraging. In this case, the encoding/decoding complexity is equal to 6.585, and the partial sum $\sum_{i=2}^{k} \lambda_i$ (or equivalently, $\sum_{i=2}^{k} \rho_i$) exceeds 0.95 for $k \geq 29$. For comparison, consider the check-regular NSIRA ensemble in [8, Theorem 2] which requires more than 300 terms so that the partial sum $\sum_{i=2}^{k} \lambda_i$ exceeds 0.95. This significant improvement in the convergence rate of the degree distributions yields codes whose performance for moderate block lengths is superior to previous constructions. The considered ensemble of self-matched ARA codes has the property that for a design rate of one-half, $L = R$ and $\lambda = \rho$, so the d.d. pairs of the punctured bits and the parity-checks coincide.

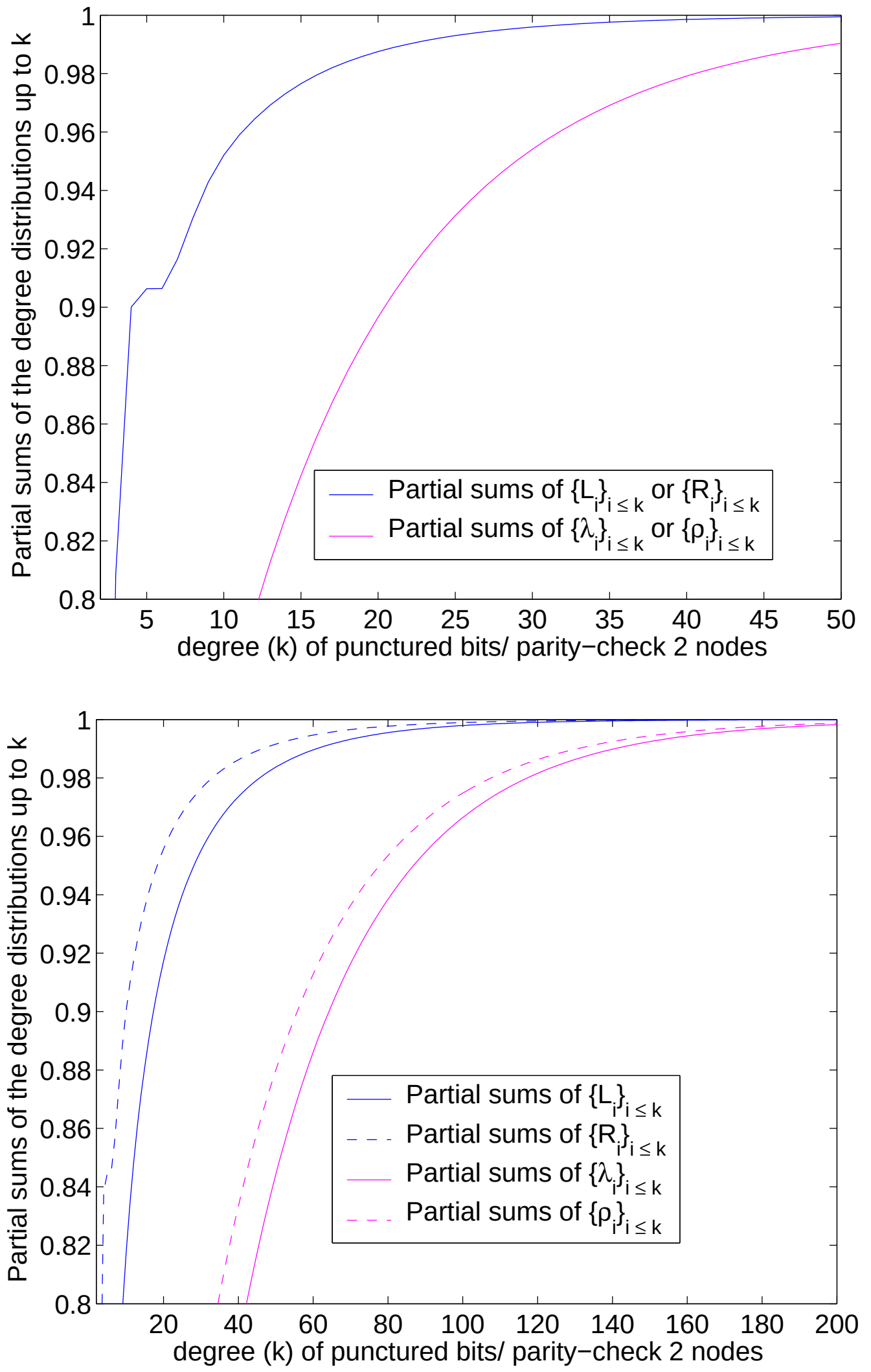

Figure 6: The plots refer to the ensemble of self-matched ARA codes with the pair of degree distributions in (45) and (46); these plots show the partial sums $\sum_{i=2}^{k} L_i$, $\sum_{i=2}^{k} R_i$, $\sum_{i=2}^{k} \lambda_i$, and $\sum_{i=2}^{k} \rho_i$ as a function of the integer $k$. The upper plot refers to $p = 0.500$ and $b = 0.9304$ (where for this case, $L_i = R_i$ and $\lambda_i = \rho_i$ for all $i \in \mathbb{N}$), and the lower plot refers to $p = 0.600$ and $b = 0.972$. The upper and lower plots refer to the (encoding and decoding) complexity per information bit which, under message-passing iterative decoding, is equal to 8.585 and 13.776, respectively.

In Fig. 7, we compare the asymptotic expressions of the degree distributions $\{L_k\}$, $\{R_k\}$, $\{\lambda_k\}$, $\{\rho_k\}$ to their exact values. There is a good match between the asymptotic and exact values for moderate to large values of $k$. The best match between the two expressions is obtained when $p = 0.50$ because this affords the minimal value of $b$. To see this phenomenon exactly, one can look at the error terms of the asymptotic expressions given in Appendix B.3.

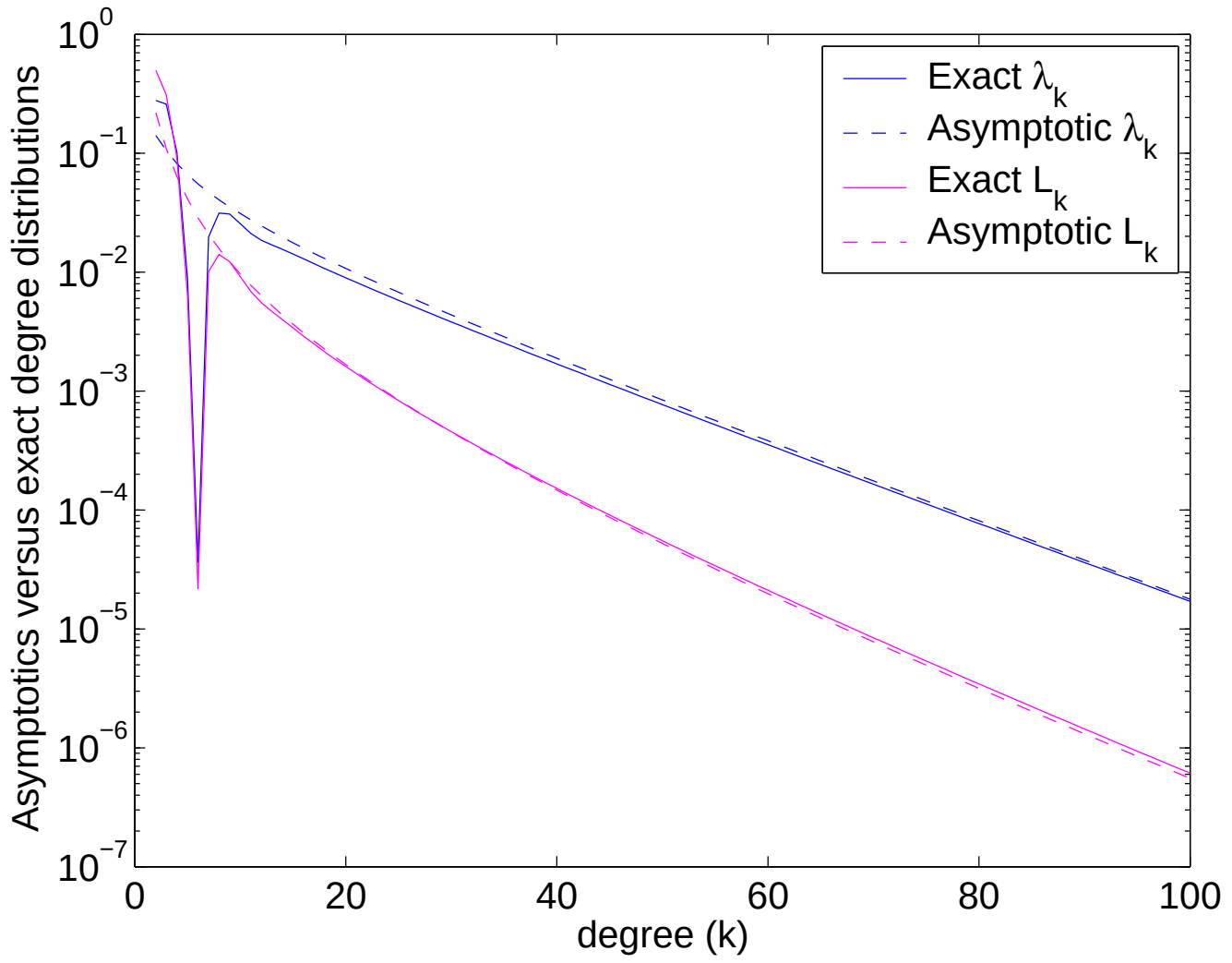

Figure 7: A plot comparing the asymptotic versus exact expressions of the degree distributions $\{L_k\}$, $\{R_k\}$, $\{\lambda_k\}$ and $\{\rho_k\}$ of the self-matched ARA ensemble, designed for a BEC with erasure probability of $p = \frac{1}{2}$. The parameter $b$ in the left and right degree distributions (45) and (46), respectively, is determined by Eq. (48), so its value is $b = 0.9304$.

To conclude, we note that the ensemble of self-matched ARA codes without puncturing, as considered in this section is well suited for moderate rates while, on the other hand, the ensembles of bit-regular and check-regular ARA codes are well suited for high and low rates, respectively. In order to make the ensemble of self-matched ARA codes suitable for high code rates, we use random puncturing (as will be exemplified later, the performance of these ensembles with puncturing is good also for moderate block lengths).

## 6.2 Further Discussion on the Construction of ARA Codes From LDPC Codes

The starting point in Section 6.1 was the choice of the function $f$ in (43) which belongs to the set $\mathcal{P}$ and which also satisfies the property $f = \mathcal{T}f$. This choice simplifies the analysis in Section 6.1 by setting $\widetilde{\lambda}(x) = \widetilde{\rho}(x) = f(x)$. From the symmetry property stated in Section 3, we see that $L$ and $R$ (and also $\lambda$ and $\rho$) have the same form except that $p$ is replaced by $1-p$. The function $f$ in (43) is however not the only function in the set $\mathcal{P}$ which satisfies the property $f = \mathcal{T}f$. In [3, Appendix V], there is a discussion on the fixed points of the operator $\mathcal{T}$. We cite here a necessary condition for the satisfaction of this property.

*Proposition* ([3, Appendix V]): If $f = \mathcal{T}f$ for some $f \in \mathcal{P}$, then $f(x) = t(x) + 1 - x$ where $t(x)$ satisfies the identity

$$t(x) = 2x - 1 - h(t(x)), \quad x \in [0,1] \tag{52}$$

and the function $h$ satisfies the properties

$$0 \le h(t) < 1, \quad h(t) = h(-t),\ t \in [0,1], \quad h(1) = h(-1) = 0.$$

Choosing appropriate functions in (52) may lead to new families of fixed points of the operator $\mathcal{T}$. For example, the choice $h(t) = \alpha(1-t^2)$ gives the function

$$f(x) = 1 + \frac{1}{2\alpha} - x - \sqrt{\left(1+\frac{1}{2\alpha}\right)^2 - \frac{2x}{\alpha}} \tag{53}$$

which indeed belongs to the set $\mathcal{P}$ and satisfies the property $f = \mathcal{T}f$ for $0 < \alpha \leq \frac{1}{2}$. Using an approach similar to the one in Section 6.1, we set $\widetilde{\lambda} = \widetilde{\rho} = f$, and rely on (41) to calculate $\widetilde{L}$ and $\widetilde{R}$. These functions are identical because $f$ is a fixed point of the operator $\mathcal{T}$. The choice of the function in (53) gives

$$\begin{aligned}\widetilde{L}(x) = \widetilde{R}(x) &= \frac{\int_0^x f(t)\,dt}{\int_0^1 f(t)\,dt} \\ &= \frac{\left(1+\frac{1}{2\alpha}\right)x - \frac{x^2}{2} + \frac{\alpha}{3}\left\{\left[\left(1+\frac{1}{2\alpha}\right)^2 - \frac{2x}{\alpha}\right]^{\frac{3}{2}} - \left(1+\frac{1}{2\alpha}\right)^3\right\}}{\frac{1}{2} - \frac{2\alpha}{3}}\end{aligned}$$

and the candidate d.d. pair $(L, R)$ is then calculated by using (42). From our numerical experiments, $L$ is likely to have a non-negative power series expansion for $p \in [0.85, 1)$ (with $\alpha \in [0.38, 0.5]$). As we noted above, the expressions for $L$ and $R$ are the same except that $p$ is replaced by $1-p$. Since the intervals $[0.85, 1)$ and $(0, 0.15]$ do not overlap, we find that both $L(x)$ and $R(x)$ cannot simultaneously have non-negative power series expansions about $x = 0$. This shows that not all self-matched functions give rise to valid degree distributions of ARA ensembles.

### 6.3 Capacity-Achieving NSIRA Ensembles from Self-Matched LDPC Codes

In this section, we construct ensembles of NSIRA codes using LDPC codes whose degree distributions from the edge perspective are matched. We apply here the concept of DE via graph reduction to ensembles of NSIRA codes. In this case, the graph reduction only applies to the "parity-check 2" nodes (see Fig. 2). This is because the upper part of Fig. 2 does not exist in the Tanner graph of NSIRA codes (i.e., the "punctured bit" nodes in this figure are the "information bit" nodes in the graph of NSIRA codes). Based on graph reduction, we obtain that $L = \widetilde{L}$ for ensembles of NSIRA codes, while the functions $R$ and $\widetilde{R}$ satisfy the equality in (42). In a similar manner, the equality $\lambda = \widetilde{\lambda}$ holds for NSIRA ensembles while equality (12) is satisfied for the degree distributions of the parity-checks from the edge perspective. We note that from (12) and (13), the fixed point of the DE equations for NSIRA ensembles is given by

$$\lambda\left(1 - \frac{(1-p)^2 \rho(1-x)}{\big(1-pR(1-x)\big)^2}\right) = x.$$

Of course, this equation coincides with the DE fixed point equation [8, Eq. (6)] (with $x_0$ replaced by $x$) derived previously for NSIRA codes.

For the construction of ensembles of NSIRA codes using LDPC codes whose degree distributions from the edge perspective are both matched to themselves, we rely as a starting point on the function $f$ in (43) which forms a d.d. which is matched to itself, and set $\widetilde{\lambda}(x) = \widetilde{\rho}(x) = \frac{(1-b)x}{1-bx}$ for $0 < b < 1$, similarly to Section 6.1. For the considered ensemble of NSIRA codes, the d.d. $L(x)$ is then equal to $\widetilde{L}(x)$ in (44), i.e.,

$$L(x) = \frac{bx + \ln(1-bx)}{b + \ln(1-b)}\,. \tag{54}$$

From this, we see that there are no degree-1 "information bit" nodes, and that the fraction of "information bit" nodes with degree-$i$ is given by

$$L_i = -\frac{b^i}{i}\,\frac{1}{b+\ln(1-b)}\ ,\quad i = 2,3,\dots\ .$$

The non-negativity of the sequence $\{L_i\}$ holds when $0 < b < 1$ (so $b + \ln(1-b) < 0$). Therefore, the power series expansion of the d.d. $L$ is always non-negative and there is no requirement on the erasure probability $p$ in this regard. The condition for the d.d. $R$ to be non-negative is identical to that of the self-matched ARA codes derived in Appendix C.3.3 and is given by

$$p \le 1 - \frac{1}{1 - \frac{13-\sqrt{61}}{9}\,[b+\ln(1-b)]}\ . \tag{55}$$

By comparing it to the parallel requirement for the ARA ensemble, as given in (47), one observes that (55) requires a weaker condition on $p$ which is only the upper bound on $p$ in (47). As mentioned above, the d.d. $R$ is the same as for the ARA ensemble in Section 6.1. The encoding and decoding complexities of this ensemble are equal and have the form

$$\chi_{\rm E} = \chi_{\rm D} = \frac{2}{1-p} - \frac{b^2}{(1-b)\,[b+\ln(1-b)]}\ . \tag{56}$$

This gives an explicit construction of NSIRA ensembles from LDPC codes whose degree distributions from the edge perspective are matched to themselves. In general, we find by computer simulations for finite-length codes over the BEC that ARA codes have the best performance.

## 6.4 Capacity-Achieving ALDPC Ensembles from Self-Matched LDPC Codes

In this section, we construct ensembles of ALDPC codes using LDPC codes whose degree distributions from the edge perspective are matched. Using the symmetry property between NSIRA and ALDPC codes, this construction follows almost trivially from the results of Section 6.3. The symmetry transformation acts by switching $\lambda$ with $\rho$ and $p$ with $1-p$, which gives

$$R(x) = \frac{bx+\ln(1-bx)}{b+\ln(1-b)}\ . \tag{57}$$

From this, we see that there are no degree-1 "parity-check 2" nodes and that the fraction of "parity-check 2" nodes with degree-$i$ is given by

$$R_i = -\frac{b^i}{i}\,\frac{1}{b+\ln(1-b)}\ ,\quad i = 2,3,\dots\ .$$

The non-negativity of the sequence $\{R_i\}$ follows directly because $0 < b < 1$ (so $b + \ln(1-b) < 0$). Since the $R_i$ are non-negative for $p \in [0,1]$, the the valid range of $p$ for this construction is determined by the non-negativity of $L(x)$. Similar to the NSIRA ensemble in Section 6.3, the d.d. $L$ is equal to the one in (45) for the ARA ensemble. From Appendix C.3.3, we see that $L(x)$ will have a non-negative power series expansion if

$$p \ge \frac{1}{1 - \frac{13-\sqrt{61}}{9}\,[b+\ln(1-b)]}\ . \tag{58}$$

Using (27) and (45), the decoding complexity of this ensemble is given by

$$\chi_{\rm D} = \frac{3}{1-p} - \frac{b^2 p}{(1-p)(1-b)[b+\ln(1-b)]}\ . \tag{59}$$

This gives an explicit construction of ALDPC ensembles from ensembles of LDPC codes with self-dual degree distributions. In Section 7, we compare the performance of this ensembles with the ensembles of NSIRA and ARA codes in Sections 6.1 and 6.3, respectively. In general, we find that the ARA codes have the best performance.

# 7 Computer Simulations

In this section, we describe the details of our computer simulations and discuss the results. In particular, our main result is that the ARA ensemble of Section 6 shows a distinct advantage over all the other c.a. ensembles we consider. We believe this advantage is largely due to the exponential decay of its d.d. tails.

## 7.1 Encoding Details

In this section, we consider the explicit construction of finite-length ARA codes from infinite degree distributions. Let $k$ be the number of "systematic bit" nodes and $n-k$ be the number of "parity-check 2" nodes (see Fig. 2 in p. 4). In the first step, we scale and quantize the degree distributions into a list of integers corresponding to the number of "punctured bit" nodes and "parity-check 2" nodes of each degree. All "parity-check 2" nodes with degree greater than $d_{\mathrm{R}}$ are reduced to degree-$d_{\mathrm{R}}$. This truncation has the effect of reducing the number of edges between "parity-check 2" nodes and "punctured bit" nodes. All "punctured bit" nodes with degree greater than $d_{\mathrm{L}}$ are converted into pilot bits. Therefore, the values of these bits are forced to a known value (e.g., zero) at the encoder. This is done by choosing the value of one of the "systematic bit" nodes carefully as described in Section 2.2. We note that each pilot bit has the effect of reducing the dimension of the code by 1.

The main reason for introducing pilot bits is that decoding cannot get started without them (see the DE equations in Section 2.3.1). A slight modification of the encoding process is required for pilot bits, and we describe it now using the notation from Section 2.2. Our goal is to set the punctured bit $v_j$ to zero by picking a systematic bit appropriately. We find that the punctured bit $v_j$ can be set to zero by choosing the systematic bit $u_j$ to be $v_{j-1} = \sum_{i=1}^{j-1} u_i$. Since each pilot bit determines one of the systematic bits, this process reduces the actual number of information bits by the number of pilot bits.

While long random codes chosen from the ensemble of Section 6.1 tend to have vanishing bit erasure probability as the block length goes to infinity, the block erasure probability does not vanish. This phenomenon is caused by small weaknesses in the graph, and can be mitigated by using a high-rate outer code as a second layer of protection. In the case of ARA codes, we believe that it is most effective to apply this code to the punctured bits. In particular, we force the last $m$ punctured bits to equal random linear combinations of the first $k-m$ punctured bits. This operation reduces the information content of the code by $m$ bits.

Again, we must slightly modify the encoding process to accommodate the outer code. Using the notation of Section 2.2, we find that this involves forcing the sequence $\underline{v}$ to satisfy $H\underline{v} = \underline{0}$ for some $m \times k$ parity-check matrix $H$. For our purposes, it suffices to consider matrices of the form $H = [P\ I]$, where $I$ is the $m \times m$ identity and $P$ is an $m \times (k-m)$ matrix with entries $P_{j,i}$. We can require the $\underline{v}$ sequence to satisfy $H\underline{v} = \underline{0}$ by choosing each $u_j$, for $j \in \{k-m+1, \ldots, k\}$, as

follows. We choose

$$u_j = v_{j-1} + \sum_{i=1}^{k-m} P_{j,i} v_i$$

because combining this with (1) shows that

$$v_j = \sum_{i=1}^{k-m} P_{j,i} v_i \qquad \Rightarrow \qquad H\underline{v} = \underline{0}.$$

## 7.2 Decoding Details

The simulations essentially use Luby's decoding algorithm [1], which starts with the full graph and deletes edges as they become known. This process starts by choosing a degree-1 parity-check node and finding the bit node to which it attaches. This bit node is declared known, and all of the edges which are attached to it are removed from the graph. Without loss of generality, one can assume that the all-zero codeword was transmitted, and the received bits are erased with probability $p$.

One advantage of Luby's decoding algorithm for LDPC codes is that it averages over all random graphs without explicitly constructing each graph. It does this simply by tracking the number of nodes of each degree throughout the decoding process. For ARA codes, we first use graph reduction (from Section 2.3.2) to convert the ARA graph into an LDPC graph. This reduction is not done in the average sense, but instead by explicitly placing erasures and combining nodes. This step implicitly averages over all orderings of the systematic and parity bits. Next, the resulting LDPC code is decoded using Luby's algorithm. This step implicitly averages over all random graphs. Therefore, this simulation technique averages over the entire ARA ensemble. Careful design of the graph can only improve performance.

If decoding terminates before all bits are known, then the high-rate outer code is decoded. Assume that $l$ bits remain unknown after iterative decoding finishes. Decoding the outer code is equivalent to solving a set of $m$ linear equations with $l$ unknowns. Of course, the parity checks leftover from iterative decoding can be used to increase the number of equations (or reduce the number of unknowns). In particular, it is easy to use one of the leftover degree-2 parity-check nodes to reduce the number of unknowns by one. Such a parity check implies the equality of the two unknowns connected to it. This equality allows one to add two columns of the matrix together and reduce the number of unknowns by one. This process can be continued, but one must be wary of linear dependencies among the degree-2 parity-check nodes. Therefore, if $t$ is the number of linearly independent degree-2 parity-check nodes remaining at the end of iterative decoding, then decoding is successful if and only if the rank of the new $m \times (l-t)$ matrix is $l-t$. We note that each entry in the $P$ matrix is chosen randomly from $\{0,1\}$ with equal probability. Choosing the $P$ matrix more carefully can only improve performance.

## 7.3 Discussion of Results

In this section, we discuss the results of the computer simulations. Although all of the ensembles introduced have been simulated, results are presented mainly for the ARA and IRA ensembles which are constructed from self-matched LDPC codes. These codes seem to have the best performance in the waterfall region (probably because their degree distributions decay exponentially fast). The results are compared with simulations of Shokrollahi's check-regular ensemble [2].

Figures 8 and 9 compare the self-matched ARA, the self-matched IRA, and the check-regular LDPC ensembles at rate $\frac{1}{2}$. Figures 8 shows the raw error rates without using a high rate outer code. Therefore, the error floor is rather severe because the results are averaged over the entire ensemble and no attempt was made to avoid small stopping sets. Figure 9 shows the results when a high rate outer code is used to mitigate small stopping sets. The results show that the self-matched ARA ensemble can handle an erasure rate roughly 0.005 larger than the other ensembles while maintaining the same performance. This gain is present throughout the waterfall region and similar for block lengths of 8192 and 65536 bits.

Fig. 10 shows how the performance of the self-matched ARA ensemble varies with block length. The upper plot shows the results without a high rate outer code and one notices that the word erasure probability never goes below 10%. Even with the high rate outer code, this ensemble has a word erasure floor due to the fraction of degree-2 bit nodes. As with stable ensembles of irregular LDPC codes, this floor can be made arbitrarily low by expurgating low weight stopping sets and/or adding a stronger outer code.

For a rate $R > \frac{1}{2}$, one can either design self-matched ARA codes directly for this rate or, alternatively, by first designing rate–$\frac{1}{2}$ self-matched ARA codes and puncturing the code bits up to rate $R$. The problem with designing the code directly for rate $R$ is that the parameter $b$ in (45) and (46) must be increased in this case and becomes very close to 1. This increases the encoding and decoding complexities and the required maximum degree (see Fig. 5). For example, the rate–$\frac{1}{2}$ ensemble requires only about the 30 first terms of the degree distributions in order to achieve 99% of the design rate while the rate–$\frac{7}{10}$ ensemble requires about the 160 first terms of these degree distributions. Figure 11 shows the performance of these two design methods for rate $\frac{7}{10}$. The results are compared directly in Fig. 12, showing the advantage of the methodology where the ensemble is designed for rate one-half and then punctured to obtain the higher rate. This advantage over the approach of designing self-matched ARA codes without puncturing is exemplified in Fig. 11 either if the ARA code is combined with a high-rate outer code or not.

It can be observed from Fig. 5 that the ensemble of self-matched ARA codes with the pair of degree distributions in (45) and (46) is not suitable for designing codes of low rates; the lower the design rate becomes below $\frac{1}{2}$ bits per channel use, the complexity of this ensemble increases significantly (see the left plot of Fig. 5). To this end, we propose the bit-regular accumulate-LDPC (ALDPC) codes (see Section 6.4) as a preferable alternative for designing codes of low rates. The performance of the bit-regular ALDPC ensemble, where the degree of the bit nodes is set to 3, is shown in Fig. 13. In this figure, the performance of these codes is exemplified for moderate to large block lengths, showing the significance of a high-rate outer code in reducing the erasure floor.

# 8 Summary and Conclusions

In this section, we provide a table of c.a. code constructions for the BEC. With the exception of the LDPC and systematic irregular repeat-accumulate (SIRA) codes, all of these codes achieve capacity on the BEC with bounded complexity per information bit. Since each of the actual c.a. degree distributions has infinite support, we let $M$ be the truncation depth of the d.d. and give the gap to capacity as a function of $M$.

| Code | Type | Range of $p$ | Bit d.d. | Check d.d. | Gap to capacity ($\varepsilon$) | Complexity as $\varepsilon \to 0$ |
|---|---|---|---|---|---|---|
| LDPC | Tornado | (0,1) | $\lambda(x) = -\frac{1}{\alpha}\ln(1-x)$ | $\rho(x) = e^{\alpha(x-1)}$ | $O\left(\frac{1}{M}\right)$ | $\chi_{\text{D}} = O\left(\log\frac{1}{\varepsilon}\right)$ |
| LDPC | CR-$k$ | $(0,1)$ | $\lambda(x) = 1-(1-x)^{\frac{1}{k-1}}$ | $\rho(x) = x^{k-1}$ | $O\left(\frac{1}{M}\right)$ | $\chi_{\text{D}} = O\left(\log\frac{1}{\varepsilon}\right)$ |
| SIRA | [5, Th. 2] | (0,1) | [5, Eq. (33)] | $\rho(x) = e^{\alpha(x-1)}$ | $O\left(\frac{1}{M}\right)$ | $\chi_{\text{E}} = \chi_{\text{D}} = O\left(\log\frac{1}{\varepsilon}\right)$ |
| NSIRA | CR-3 | $(0, 0.95)$ | [8, Eq. (15)] | $\rho(x) = x^2$ | $O\left(\frac{1}{M^{\frac{3}{2}}}\right)$ | $\chi_{\text{E}} = \chi_{\text{D}} = \frac{5}{1-p}$ |
| NSIRA | BR-3 | $(0, \frac{1}{13})$ | $\lambda(x) = x^2$ | [8, Eq. (10)] | $O\left(\frac{1}{\sqrt{M}}\right)$ | $\chi_{\text{E}} = \chi_{\text{D}} = 3 + \frac{2}{1-p}$ |
| NSIRA | SM | Eq. (55) | Eq. (54) | Eq. (46) | $O\left(\frac{b^M}{\ln^2(M)}\right)$ | Eq. (56) |
| ARA | CR-3 | $(0.616, 1)^\dagger$ | Eq. (37) | $R(x) = x^3$ | $O\left(\frac{1}{M^{\frac{3}{2}}}\right)$ | $\chi_{\text{E}} = \chi_{\text{D}} = 3 + \frac{5p}{1-p}$ |
| ARA | BR-3 | $(0, 0.348)$ | $L(x) = x^3$ | Eq. (36) | $O\left(\frac{1}{M^{\frac{3}{2}}}\right)$ | $\chi_{\text{E}} = \chi_{\text{D}} = 6 + \frac{2p}{1-p}$ |
| ARA | SM | Eq. (47) | Eq. (45) | Eq. (46) | $O\left(\frac{b^M}{\ln^2(M)}\right)$ | Eq. (49) |
| ALDPC | CR-3 | $(\frac{12}{13}, 1)^\dagger$ | Eq. (40) | $\rho(x) = x^2$ | $O\left(\frac{1}{\sqrt{M}}\right)$ | $\chi_{\text{D}} = \frac{3(1+p)}{1-p}$ |
| ALDPC | BR-3 | $(0.05, 1)$ | $\lambda(x) = x^2$ | Eq. (39) | $O\left(\frac{1}{M^{\frac{3}{2}}}\right)$ | $\chi_{\text{D}} = \frac{6}{1-p}$ |
| ALDPC | SM | Eq. $(58)^\dagger$ | Eq. (45) | Eq. (57) | $O\left(\frac{b^M}{\ln^2(M)}\right)$ | Eq. (59) |

Table 1: List of capacity-achieving (c.a.) codes for the BEC. In this table, 'BR', 'CR' and 'SM' stand for 'bit-regular', 'check-regular' and 'self-matched degree distributions', respectively. The valid range of $p$ is marked by † if it can be extended to $(0, 1)$ via puncturing. The parameter $b$ in the d.d. pairs of the SM ensembles is allowed to be in the interval $(0.9304, 1)$.

Simulation results show that among all these ensembles, the self-matched ARA ensemble has a superior performance for moderate to large block lengths (considering rates which are at least $\frac{1}{2}$ bits per channel use). We believe the performance advantage of the self-matched ARA ensemble is mainly due to the *exponential* decay of the d.d. coefficients. For designing low-rate codes, we propose the bit-regular accumulate-LDPC (ALDPC) codes in Section 6.4 as a preferable alternative.

# Acknowledgment

This research work was supported by a grant from Intel Israel. The work of I. Sason was also supported by the Taub and Shalom Foundations. The authors acknowledge Rüdiger Urbanke for stimulating discussions.

# Appendix A
# Derivation of $Q$ in (35)

Following the procedure of Section 5.1, and starting from (30) gives

$$\begin{aligned} Q(x) &= \frac{\int_0^x \widetilde{\rho}(t)\,dt}{\int_0^1 \widetilde{\rho}(t)\,dt} \\ &= \frac{3}{p}\int_0^x \widetilde{\rho}(t)\,dt \end{aligned} \tag{A.1}$$

where the last equality follows from (21) and (22), i.e.,

$$\int_0^1 \widetilde{\rho}(t)\,dt = \int_0^1 \widetilde{\lambda}(t)\,dt = \frac{p}{L'(1)} = \frac{p}{3}.$$

The substitution $t = \widetilde{\rho}^{-1}(u)$ and integration by parts gives

$$\begin{aligned} &\int_0^x \widetilde{\rho}(t)\,dt \\ &= \int_0^{\widetilde{\rho}(x)} u\frac{d}{du}\left(\widetilde{\rho}^{-1}(u)\right)du \\ &= u\widetilde{\rho}^{-1}(u)\Big|_{u=0}^{\widetilde{\rho}(x)} - \int_0^{\widetilde{\rho}(x)} \widetilde{\rho}^{-1}(u)du \\ &= x\widetilde{\rho}(x) - \int_0^{\widetilde{\rho}(x)} \left[1 - \frac{p^2(1-u)^2}{\left(1-(1-p)(1-u)^3\right)^2}\right]du \\ &= (x-1)\widetilde{\rho}(x) - \frac{p^2}{3(1-p)}\frac{1}{1-(1-p)(1-u)^3}\Big|_{u=0}^{\widetilde{\rho}(x)} \\ &= (x-1)\widetilde{\rho}(x) + \frac{p}{3}\frac{1-\left(1-\widetilde{\rho}(x)\right)^3}{1-(1-p)\left(1-\widetilde{\rho}(x)\right)} \end{aligned}$$

and the substitution of the last equality in the RHS of (A.1) gives (35).

# Appendix B
# Capacity-Achieving Self-Matched Ensembles of ARA Codes

From (45), we get

$$\begin{aligned}
L(x) &= \frac{1}{1-p}\left(1 - \frac{1}{1+\frac{(1-p)\left[bx+\ln(1-bx)\right]}{p\left[b+\ln(1-b)\right]}}\right) \\
&= \frac{1}{1-p}\left(1 - \frac{1}{1-\alpha(p,b)\left[bx+\ln(1-bx)\right]}\right) \\
&= -\frac{1}{1-p}\sum_{m=1}^{\infty} \alpha^m(p,b)\ \left[bx+\ln(1-bx)\right]^m \\
&= \frac{1}{1-p}\sum_{m=1}^{\infty}\left\{(-1)^{m-1}\ \alpha^m(p,b)\left[\frac{b^2x^2}{2}+\frac{b^3x^3}{3}+\frac{b^4x^4}{4}+\ldots\right]^m\right\} \qquad \text{(B.1)}
\end{aligned}$$

where

$$\alpha(p,b) \triangleq -\frac{1-p}{p\left[b+\ln(1-b)\right]}\ . \qquad \text{(B.2)}$$

We note that $\alpha(p,b)$ is positive for $0 < b < 1$ and $0 < p < 1$. From equation (B.1), it is easily verified that the minimal degree in the power series expansion of $\left[bx+\ln(1-bx)\right]^m$ is $x^{2m}$, so if $k < 2m$, then the coefficient of $x^k$ is equal to zero. Since $m$ and $k$ are integers, then if follows that the coefficient of $x^k$ in the power series expansion of $\left[bx+\ln(1-bx)\right]^m$ vanishes for $m \geq \lfloor \frac{k}{2} \rfloor + 1$. This gives the following equality:

$$L_k = [x^k]L(x) = -\frac{1}{1-p}\sum_{m=1}^{\lfloor \frac{k}{2}\rfloor} \alpha^m(p,b)\cdot [x^k]\left\{\left[bx+\ln(1-bx)\right]^m\right\}$$

where the infinite sum we had before turned to be a finite sum in the last equality. Therefore, the power series expansion of $bx + \ln(1-bx)$ yields that

$$\begin{aligned}
L_k &= \frac{1}{1-p}\sum_{m=1}^{\lfloor \frac{k}{2}\rfloor}\left\{(-1)^{m-1}\ \alpha^m(p,b)\cdot\left\{[x^k]\left(\frac{b^2x^2}{2}+\frac{b^3x^3}{3}+\frac{b^4x^4}{4}\ldots\right)^m\right\}\right\} \\
&= \frac{\alpha(p,b)}{1-p}\sum_{m=1}^{\lfloor \frac{k}{2}\rfloor}\left\{(-1)^{m-1}\ \alpha^{m-1}(p,b)\cdot\left\{[x^k]\left(\frac{b^2x^2}{2}+\frac{b^3x^3}{3}+\frac{b^4x^4}{4}\ldots\right)^m\right\}\right\}
\end{aligned}$$

From the last equality, we obtain that

$$L_k = \frac{\alpha(p,b)\ b^k}{1-p}\sum_{m=1}^{\lfloor \frac{k}{2}\rfloor}\left\{(-1)^{m-1}\ c_{m,k}\cdot\alpha^{m-1}(p,b)\right\}\ , \qquad k = 2,3,4,\ldots \qquad \text{(B.3)}$$

where

$$c_{m,k} \triangleq \sum_{\substack{i_1+i_2+\ldots+i_m=k \\ i_1,i_2,\ldots,i_m\geq 2}} \frac{N_{i_1,i_2,\ldots,i_m}}{i_1\ i_2\ \ldots\ i_m}\ , \qquad m=1,2,\ldots,\lfloor\frac{k}{2}\rfloor \tag{B.4}$$

and $N_{i_1,i_2,\ldots,i_m}$ in the RHS of (B.4) designates the number of different permutations of the sequence $\{i_1,i_2,\ldots,i_m\}$. From the duality between $R$ and $L$ in our example (see (45) and (46)), then for the calculation of the coefficient $R_k$ in the power series expansion of $R$ we only need to replace $p$ in (B.3) by $1-p$, so

$$R_k = \frac{\alpha(1-p,b)\ b^k}{p} \sum_{m=1}^{\lfloor\frac{k}{2}\rfloor} \Big\{(-1)^{m-1}\ c_{m,k}\cdot\alpha^{m-1}(1-p,b)\Big\}\ , \qquad k=2,3,4,\ldots\ . \tag{B.5}$$

Since $\lambda(x)=\frac{L'(x)}{L'(1)}$, then $\lambda_k=\frac{kL_k}{L'(1)}$. From (B.2) and (49), we obtain the equality $L'(1)=\frac{b^2p^2\alpha(p,b)}{(1-p)(1-b)}$, and then it follows from (B.3) and the last two equalities that

$$\lambda_k = \frac{(1-b)\ k\ b^{k-2}}{p^2} \sum_{m=1}^{\lfloor\frac{k}{2}\rfloor} \Big\{(-1)^{m-1}\ c_{m,k}\cdot\alpha^{m-1}(p,b)\Big\}\ , \qquad k=2,3,4,\ldots\ . \tag{B.6}$$

From the duality between (45) and (46), then by switching $p$ in (B.6) by $1-p$, we obtain that

$$\rho_k = \frac{(1-b)\ k\ b^{k-2}}{(1-p)^2} \sum_{m=1}^{\lfloor\frac{k}{2}\rfloor} \Big\{(-1)^{m-1}\ c_{m,k}\cdot\alpha^{m-1}(1-p,b)\Big\}\ , \qquad k=2,3,4,\ldots \tag{B.7}$$

so (B.6) and (B.7) provide explicit expressions for the coefficients of the d.d. pair from the edge perspective.

### B.1 A Recursion for the Sequence $\{c_{m,k}\}$

We intend to give now an efficient way to calculate the coefficients $c_{m,k}$ where $1\leq m\leq\lfloor\frac{k}{2}\rfloor$ (as otherwise $c_{m,k}$ is equal to zero). To this end, we can rewrite $c_{m,k}$ in (B.4) as

$$c_{m,k} \triangleq \sum_{\substack{i_1+i_2+\ldots+i_m=k \\ i_1,i_2,\ldots,i_m\geq 2}} \frac{1}{i_1\ i_2\ \ldots\ i_m}\ , \qquad m=1,2,\ldots,\lfloor\frac{k}{2}\rfloor \tag{B.8}$$

where the difference from (B.4) is that now the order of the terms in the sequence $\{i_1,i_2,\ldots,i_m\}$ is relevant, so effectively every term of the form $\frac{1}{i_1\ i_2\ \ldots\ i_m}$ is counted $N_{i_1,i_2,\ldots,i_m}$ times in the sum above. We will show that this sequence satisfies a simple recursive equation. First, it is trivial that

$$c_{1,k}=\frac{1}{k}\ , \quad k=2,3,\ldots \tag{B.9}$$

Since we impose the constraint $i_1 + i_2 + \ldots i_m = k$ on the calculation of $c_{m,k}$, then we can rewrite $c_{m,k}$ in (B.8) as

$$\begin{aligned} c_{m,k} &= \frac{1}{k} \sum_{\substack{i_1+i_2+\ldots+i_m=k \\ i_1,i_2,\ldots,i_m \geq 2}} \frac{i_1+i_2+\ldots+i_m}{i_1\, i_2\, \ldots\, i_m} \\ &= \frac{1}{k} \sum_{\substack{i_1+i_2+\ldots+i_m=k \\ i_1,i_2,\ldots,i_m \geq 2}} \left( \frac{1}{i_2\, i_3\, \ldots i_m} + \frac{1}{i_1\, i_3\, \ldots\, i_m} + \ldots + \frac{1}{i_1\, i_2\, i_3\, \ldots\, i_{m-1}} \right) \end{aligned}$$

Symmetry considerations imply that the sum w.r.t. every term among the $m$ terms above are all equal to each other, so the expression of $c_{m,k}$ can be simplified to

$$c_{m,k} = \frac{m}{k} \sum_{\substack{i_1+i_2+\ldots+i_m=k \\ i_1,i_2,i_3\ldots,i_m \geq 2}} \frac{1}{i_2\, i_3\, \ldots\, i_m}\,. \tag{B.10}$$

Comparing to (B.8), the last summation is over all the possible vectors $\{i_2, i_3, \ldots, i_m\}$ (i.e., $i_1$ does not appear in the term $\frac{1}{i_2\, i_3\, \ldots i_m}$, but only $i_2,\ i_3,\ \ldots\ , i_m$ appear there). Since $i_j \geq 2$ for $1 \leq j \leq m$, then the sum $i_2+i_3+\ldots i_m$ gets all possible values between $2(m-1)$ and $k-2$ (where the inequality $2(m-1) \leq k-2$ is automatically satisfied because of the assumption that $k \geq 2m$, as otherwise $c_{m,k} = 0$). Since the summations in (B.8) and (B.10) are taken w.r.t. all possible combinations of $\{i_1, i_2, \ldots, i_m\}$ when the two constraints in (B.10) are satisfied, then we obtain the following recursive equation:

$$c_{m,k} = \frac{m}{k} \sum_{j=2(m-1)}^{k-2} c_{m-1,j}\,. \tag{B.11}$$

The combination of the recursive equation in (B.11) with the initial values in (B.9) gives an efficient way to calculate the terms of the sequence $\{c_{m,k}\}$. We implemented this algorithm in software.

## B.2 An Algorithm for Calculating the Degree Distributions in Section 6.1

We provide here an algorithm to calculate the coefficients in the d.d. pairs $(L, R)$ and $(\lambda, \rho)$. This algorithm was implemented in software.

**1.** Calculate the sequence $\{c_{m,k}\}$ with the recursive equation (B.11) and the initial values in (B.9).

**2.** Calculate $\alpha(p, b)$ from (B.2).

**3.** Calculate $\{L_k\}$ and $\{\lambda_k\}$ from (B.3) and (B.6), respectively.

**4.** Calculate $\alpha(1-p, b)$ from (B.2), and $\{R_k\}$ and $\{\rho_k\}$ from (B.5) and (B.7), respectively.

## B.3 Asymptotic Expressions for the Degree Distributions in Section 6.1

We wish to find here asymptotic expressions for the sequences $\{L_k\}$, $\{R_k\}$, $\{\lambda_k\}$ and $\{\rho_k\}$ where $k$ is sufficiently large. First we see that it is enough to solve the problem for the sequence $\{L_k\}$, since if there exists a sequence of functions $\{g_k(\cdot, \cdot)\}$ so that $L_k \approx g_k(p, b)$ for $k \gg 1$, then we obtain from (B.3)–(B.7) that the following asymptotic expressions are valid for the other degree distributions when $k \gg 1$:

$$R_k \approx g_k(1-p,b), \qquad \lambda_k \approx \frac{(1-b)k}{b^2p^2} \cdot g_k(p,b), \qquad \rho_k \approx \frac{(1-b)k}{b^2(1-p)^2} \cdot g_k(1-p,b)\,. \tag{B.12}$$

We shall therefore focus on the derivation of the asymptotic behavior of the sequence $\{L_k\}$ in the power series expansion of the function $L(\cdot)$ in (45). To this end, we rewrite $L(\cdot)$ in the form

$$\begin{aligned} L(x) &= \frac{bx+\ln(1-bx)}{p\,[b+\ln(1-b)]+(1-p)\,[bx+\ln(1-bx)]} \\ &= \frac{1}{1-p}\left(1-\frac{p[b+\ln(1-b)]}{p\,[b+\ln(1-b)]+(1-p)\,[bx+\ln(1-bx)]}\right) \\ &= \frac{1}{1-p}\left(1-\frac{1}{1+\frac{1-p}{p[b+\ln(1-b)]}\cdot[bx+\ln(1-bx)]}\right) \\ &= \frac{1}{1-p}\left(1-\frac{1}{1-\alpha(p,b)\cdot[bx+\ln(1-bx)]}\right) \end{aligned}$$

where the function $\alpha$ is introduced in (B.2), and is positive for $0<p<1$ and $0<b<1$. Let us define the function

$$h(z;c) = \frac{1}{1-c[z+\ln(1-z)]}\,, \qquad 0<c\le\frac{13-\sqrt{61}}{9}\approx 0.5766. \tag{B.13}$$

then

$$L(x) = \frac{1}{1-p}\cdot\Big(1-h\big(bx;\alpha(p,b)\big)\Big) \tag{B.14}$$

where the restriction on $c$ follows from the restriction on $\alpha$ that we obtained earlier so that $L_6$ is non-negative, and from the equality $c=\alpha(p,b)$ which holds by comparing (B.13) and (B.14).

The following equality therefore holds for $k \geq 2$:

$$L_k = -\frac{b^k}{1-p}\cdot h_k(c), \quad c \triangleq \alpha(p,b) \tag{B.15}$$

where $\{h_k(c)\}_{k\geq 0}$ is the sequence the coefficients in the power series expansion $h(z;c)=\sum_{k=0}^{\infty} h_k(c)z^k$. In the continuation, we will find the asymptotic behavior of the power series expansion of the function $h(z;c)$ in (B.13), and then use (B.15) to derive the asymptotic behavior of the sequence $\{L_k\}$, and use (B.12) for the derivation of the asymptotic behavior of the other degree distributions.

In [20], a class of methods is presented which enables one to translate, on a term-by-term basis, an asymptotic expression of a function around a dominant singularity into a corresponding asymptotic expansion for the Taylor coefficients of the function. In the continuation of the asymptotic analysis, we rely on [20]. The function $h(z;c)$ has singularities at the points where the following equation is satisfied:

$$1-c\big(z+\ln(1-z)\big)=0.$$

The closed form solution of the above equation is $z=z_{1,2}$ where

$$z_1 = 1+W(-e^{\frac{1}{c}-1})\,, \quad z_2 = z_1^*$$

and $W$ denotes the Lambert W-function. Since we require that $0<c\le\frac{13-\sqrt{61}}{9}\approx 0.5766$, then $z_1$ and $z_2$ have an absolute value which is at least equal to 2.074 (we note that the absolute value of $z_{1,2}$ is a monotonic decreasing function of $c$, so it is achieved when $c$ gets its maximal value within this interval). The dominant singularity of the function $h(z;c)$ is therefore at the point $z=1$ where the logarithm in (B.13) becomes singular. In the region around the dominant singularity at $z=1$, the function $h(z;c)$ behaves like

$$h(z;c) = O\left(\ln^{-1}\left(\frac{1}{1-z}\right)\right)$$

so from [20], the asymptotic behavior of the coefficients in the power series expansion of this function is

$$h_k = O\left(\frac{1}{k\ln^2(k)}\right).$$

A more careful analysis which is based on the singularity analysis in [20] shows that

$$h_k = \frac{1}{1-c}\,\frac{1}{k\big(1+d\ln(k)\big)^2}\left[-1+\frac{2\gamma}{1+d\ln(k)}+\frac{\pi^2-6\gamma^2}{2\big(1+d\ln(k)\big)^2}\right]+O\left(\frac{1}{k\ln^5(k)}\right) \quad \text{(B.16)}$$

where $d \triangleq \frac{c}{1-c}$, and $\gamma \approx 0.5772$ designates Euler's constant. From (B.15) and (B.16), we obtain that

$$L_k = \frac{1}{\big(1-\alpha(p,b)\big)(1-p)}\cdot\frac{b^k}{k}\cdot\frac{1}{\big(1+d(p,b)\ln(k)\big)^2}\cdot\left(1-\frac{2\gamma}{1+d(p,b)\ln(k)}\right)+O\left(\frac{b^k}{k\ln^4(k)}\right) \quad \text{(B.17)}$$

where the transition from (B.16) to (B.17) follows by substituting $c=\alpha(p,b)$ where $\alpha(\cdot,\cdot)$ is introduced in (B.2), and then based on (B.2) and (B.16)

$$d(p,b) = \frac{\alpha(p,b)}{1-\alpha(p,b)} = -\frac{1-p}{1-p+p[b+\ln(1-b)]}.$$

The asymptotic expression for the other degree distributions (i.e., $\{R_k\}$, $\{\lambda_k\}$ and $\{\rho_k\}$ follow now immediately from the asymptotic behavior of $\{L_k\}$ in (B.17) and the transition to the asymptotic behavior of the other d.d. in (B.12). The following degree distributions are therefore obtained:

$$R_k = \frac{1}{\big(1-\alpha(1-p,b)\big)p}\cdot\frac{b^k}{k}\cdot\frac{1}{\big(1+d(1-p,b)\ln(k)\big)^2}\cdot\left(1-\frac{2\gamma}{1+d(1-p,b)\ln(k)}\right)+O\left(\frac{b^k}{k\ln^4(k)}\right)$$

$$\lambda_k = \frac{1-b}{\big(1-\alpha(p,b)\big)p^2(1-p)}\cdot\frac{b^{k-2}}{\big(1+d(p,b)\ln(k)\big)^2}\cdot\left(1-\frac{2\gamma}{1+d(p,b)\ln(k)}\right)+O\left(\frac{b^k}{\ln^4(k)}\right)$$

$$\rho_k = \frac{1-b}{\big(1-\alpha(1-p,b)\big)(1-p)^2p}\cdot\frac{b^{k-2}}{\big(1+d(1-p,b)\ln(k)\big)^2}\cdot\left(1-\frac{2\gamma}{1+d(1-p,b)\ln(k)}\right)+O\left(\frac{b^k}{\ln^4(k)}\right).$$

Therefore, the asymptotic behavior of the d.d. pairs w.r.t. the nodes and the edges is given by

$$L_k, R_k = O\left(\frac{b^k}{k\ln^2(k)}\right), \qquad \lambda_k, \rho_k = O\left(\frac{b^k}{\ln^2(k)}\right).$$

The parameter $b$ above is chosen according to (48), so it is determined as a function of the erasure probability ($p$) of the BEC ($0.9304 \le b < 1$). The closer is the value of $p$ to one-half, then the smaller becomes the value of $b$, and this accelerates the convergence rate to zero of the d.d. pairs above. It is therefore clear from these two equations that the convergence rate of the d.d. pairs is improved when $p$ becomes closer to one-half (as is also shown in Fig. 6).

# Appendix C
# A Generalization of Polya's Criterion to Discrete Distributions and Its Application to Non-Negativity Proofs

## C.1 Polya's Criterion

Polya's Criterion [21, p. 509] is a simple condition which is sufficient to imply that a function $F(t)$ is the inverse Fourier transform (i.e., a characteristic function) of a non-negative function $f(x)$. It requires that the function $F$ is real, symmetric, non-negative, and convex non-increasing on $[0,\infty)$. This gives a simple method of proving that some functions are indeed characteristic functions.

**Example C.1.** Let $\{X_i\}_{i\in\mathbb{N}}$ be a sequence of random variables with characteristic function

$$F(t) = e^{-|t|^\alpha}, \quad \alpha \in (0,1].$$

Since there is no closed-form expression for the Fourier transform

$$f(x) = \frac{1}{2\pi}\int_{-\infty}^{\infty} F(t)e^{-itx}dt,$$

we cannot verify directly that $f(x) \geq 0$. It is easy, however, to verify that $F(t)$ satisfies Polya's Criterion for $\alpha \in (0,1]$ and this implies that each $X_i$ is a well-defined random variable.

## C.2 A Generalization of Polya's Criterion to Discrete Distributions

In this section, we generalize Polya's Criterion to discrete distributions and offer an elementary proof which is substantially different from the standard "tent function" proof outlined in [21, p. 505].

**Definition C.1.** A function $\gamma(x)$ has *cosine symmetry* if

$$\gamma(x) = \gamma(2\pi + x), \quad \forall\, x \in \mathbb{R}$$

and

$$\gamma(x) = -\gamma(\pi - x) = -\gamma(\pi + x) = \gamma(2\pi - x) \geq 0, \quad \forall\, x \in [0, \pi/2].$$

**Lemma C.1.** Let $\gamma(x)$ be a function with cosine symmetry which is non-negative on $[0, \frac{\pi}{2}]$, and $f(x)$ be a real function convex for $x \in [0, 2\pi)$. Then, the integral $\int_0^{2\pi}\gamma(x)f(x)dx$ is non-negative.

*Proof.* First, rewrite the integral as

$$\begin{aligned}
\int_0^{2\pi}\gamma(x)f(x)dx &= \int_0^{\pi/2}\left[\gamma(x)f(x) + \gamma(\pi-x)f(\pi-x) + \gamma(\pi+x)f(\pi+x) + \gamma(2\pi-x)f(2\pi-x)\right]dx \\
&= \int_0^{\pi/2}\gamma(x)\left[f(x) - f(\pi-x) - f(\pi+x) + f(2\pi-x)\right]dx \\
&= \int_0^{\pi/2}\gamma(x)g(x)dx
\end{aligned}$$

where $g(x) \triangleq f(x) - f(\pi - x) - f(\pi + x) + f(2\pi - x)$. Taking the derivative of the function $g$ gives

$$\begin{aligned} g'(x) &= f'(x) + f'(\pi - x) - f'(\pi + x) - f'(2\pi - x) \\ &= \left[f'(x) - f'(\pi + x)\right] + \left[f'(\pi - x) - f'(2\pi - x)\right], \end{aligned}$$

which is non-positive for $x \in \left[0, \frac{\pi}{2}\right]$ because $f''(x) \geq 0$ for $x \in [0, 2\pi]$ (i.e., this gives for $x \in [0, \frac{\pi}{2}]$ that $f'(x) \leq f'(x + \pi)$ and $f'(\pi - x) \leq f'(2\pi - x)$). Since $g\left(\frac{\pi}{2}\right) = 0$, this implies that $g(x) \geq 0$ for $x \in \left[0, \frac{\pi}{2}\right]$. Finally, we note that the integral is non-negative because both $\gamma(x)$ and $g(x)$ are non-negative for $x \in \left[0, \frac{\pi}{2}\right]$. □

**Lemma C.2.** Let $\gamma(x)$ be a function with cosine symmetry which is non-negative on $[0, \frac{\pi}{2}]$, and let $F(t)$ be a real symmetric function convex for $t \in [0, a]$. Then, for $x > 0$

$$\frac{1}{\pi} \int_0^a F(t)\gamma(tx)dt \geq \frac{1}{\pi} \int_{\frac{2\pi}{x}\left\lfloor \frac{xa}{2\pi} \right\rfloor}^a F(t)\gamma(tx)dt.$$

*Proof.* Using Lemma C.1, we find that the integral over each full cycle of $\gamma(x)$ is non-negative which leaves only the remaining partial cycle near $a$. We show this using, for $x > 0$, the decomposition

$$\begin{aligned} \frac{1}{\pi} \int_0^a F(t)\gamma(tx)dt &= \frac{1}{\pi} \int_{\frac{2\pi}{x}\left\lfloor \frac{xa}{2\pi} \right\rfloor}^a F(t)\gamma(tx)dt + \frac{1}{\pi} \sum_{j=0}^{\left\lfloor \frac{xa}{2\pi} \right\rfloor - 1} \left[ \int_{\frac{2\pi j}{x}}^{\frac{2\pi(j+1)}{x}} F(t)\gamma(tx)dt \right] \\ &= \frac{1}{\pi} \int_{\frac{2\pi}{x}\left\lfloor \frac{xa}{2\pi} \right\rfloor}^a F(t)\gamma(tx)dt + \frac{1}{\pi x} \sum_{j=0}^{\left\lfloor \frac{xa}{2\pi} \right\rfloor - 1} \left[ \int_0^{2\pi} F\left(\frac{2\pi j + u}{x}\right)\gamma(u)du \right] \\ &\geq \frac{1}{\pi} \int_{\frac{2\pi}{x}\left\lfloor \frac{xa}{2\pi} \right\rfloor}^a F(t)\gamma(tx)dt. \end{aligned}$$

The last step follows from Lemma C.1 using the fact that $F\left(\frac{2\pi j + u}{x}\right)$ is convex for $u \in [0, 2\pi]$ if $0 \leq j \leq \left\lfloor \frac{xa}{2\pi} \right\rfloor - 1$. □

**Definition C.2.** A function $F(t) : \mathbb{R} \to \mathbb{R}$ satisfies Polya's Criterion for $t \in [0, a]$, if and only if it satisfies the following conditions:

**1.** $F(t) = F(-t)$, i.e., $F$ is a symmetric function.

**2.** $F(t) \geq 0$ i.e., $F$ is non-negative over the interval $[0, a]$.

**3.** $F'(t) \leq 0$, i.e., $F$ is non-increasing over the interval $[0, a]$.

**4.** $F''(t) \geq 0$, i.e., $F$ is convex over the interval $[0, a]$.

Let $\mathcal{P}_a$ be the set of functions satisfying Polya's Criterion on $[0, a]$.

**Theorem C.2.** Any function $F \in \mathcal{P}_\infty$ is the characteristic function of a continuous probability distribution, and any function $F \in \mathcal{P}_\pi$ which is $2\pi$-periodic (i.e., $F(x) = F(x + 2\pi)$) is the characteristic function of a discrete probability distribution.

*Proof.* For the $F \in \mathcal{P}_\infty$ case, we have

$$\begin{aligned} f(x) &= \frac{1}{2\pi} \int_{-\infty}^{\infty} F(t)e^{-itx}dt \\ &= \frac{1}{\pi} \int_0^{\infty} F(t)\cos(tx)dt. \end{aligned}$$

First, we note that $f(0) \geq 0$ because $F(t) \geq 0$. Next, we apply Lemma C.2 to this with $\gamma(x) = \cos(x)$, which gives

$$\begin{aligned} f(x) &\geq \lim_{a\to\infty} \frac{1}{\pi} \int_{\frac{2\pi}{x}\left\lfloor \frac{xa}{2\pi}\right\rfloor}^{a} F(t)\cos(tx)dt \\ &\geq \frac{1}{\pi} \lim_{a\to\infty} \left(a - \frac{2\pi}{x}\left\lfloor \frac{xa}{2\pi}\right\rfloor\right) \min_{\frac{2\pi}{x}\left\lfloor \frac{xa}{2\pi}\right\rfloor \leq t \leq a} F(t)\cos(tx) \\ &\geq -\lim_{a\to\infty} \frac{2}{x} F\left(\frac{2\pi}{x}\left\lfloor \frac{xa}{2\pi}\right\rfloor\right) \end{aligned}$$

where the last step follows since the function $F \in \mathcal{P}_\infty$ is non-negative and monotonic non-increasing, so for any $a > 0$ and $x > 0$

$$|F(t)\cos(tx)| \leq F(t) \leq F\left(\frac{2\pi}{x}\left\lfloor \frac{xa}{2\pi}\right\rfloor\right) \;, \quad \forall\, t \in \left[\frac{2\pi}{x}\left\lfloor \frac{xa}{2\pi}\right\rfloor, a\right].$$

If $\lim_{a\to\infty} F(a) = 0$, then we have $f(x) \geq 0$ for $x \neq 0$. On the other hand, if $\lim_{a\to\infty} F(a) > 0$, then $f(x)$ has a Delta function at $x = 0$. Subtracting this Delta function returns us to the first case and shows that $f(x) \geq 0$ for $x \neq 0$.

For the $2\pi$-periodic $F \in \mathcal{P}_\pi$ case, we have

$$\begin{aligned} h_k &= \frac{1}{2\pi}\int_{-\pi}^{\pi} F(t)e^{-ikt}dt \\ &= \frac{1}{\pi}\int_{0}^{\pi} F(t)\cos(kt)dt. \end{aligned}$$

For $k = 0$, we see that $h_0 \geq 0$ by the non-negativity of $F(t)$ for $t \in [0, \pi]$. For $k > 0$, we apply Lemma C.2 with $\gamma(x) = \cos(x)$ to get

$$h_k \geq \frac{1}{\pi}\int_{\frac{2\pi}{k}\left\lfloor \frac{k}{2}\right\rfloor}^{\pi} F(t)\cos(kt)dt.$$

When $k$ is even, we have $h_k \geq 0$ because the range of integration becomes zero. When $k$ is odd, we have

$$\begin{aligned} h_k &\geq \frac{1}{\pi}\int_{\frac{\pi(k-1)}{k}}^{\frac{\pi(k-1/2)}{k}} F(t)\cos(tk)dt + \frac{1}{\pi}\int_{\frac{\pi(k-1/2)}{k}}^{\pi} F(t)\cos(kt)dt \\ &= \frac{1}{\pi}\int_{0}^{\frac{\pi}{2k}} \left[F\left(\frac{\pi(k-1)}{k} + u\right) - F(\pi - u)\right]\cos(ku)du. \end{aligned}$$

This gives $h_k \geq 0$ because $\cos(tk) \geq 0$ for $t \in \left[0, \frac{\pi}{2k}\right]$ and $F(t)$ is non-increasing for $t \in [0, \pi]$. □

**Corollary C.1.** Let the function $g : \mathbb{C} \to \mathbb{C}$ be analytic on the unit disc except possibly $x = 1$. If the real function $h(x) \triangleq \mathrm{Re}\left\{g\left(e^{ix}\right)\right\}$ is symmetric, convex on $[0, \pi]$, and satisfies $\int_0^\pi h(x)dx \geq 0$, then the function $g$ has a power series expansion about zero with non-negative coefficients.

*Proof.* By the analyticity of $g$, we have the power series expansion $g(x) = \sum_{n=0}^{\infty} g_n x^n$. Straightforward algebra shows that

$$h(x) = \sum_{n=0}^{\infty}\left\{\mathrm{Re}(g_n)\cos(nx) - \mathrm{Im}(g_n)\sin(nx)\right\}.$$

Since $h$ is assumed to be symmetric, then the imaginary part of $g_n$ must be zero for all non-negative integers $n$. Hence, $g(x)$ is real for $x \in [-1, 1)$. Adding the additional assumption that $h$ is convex on the interval $[0, \pi]$, implies that $h$ is monotonic non-increasing in this interval. To show this, we calculate

$$h'(x) = \text{Re}\left\{ie^{ix}g'(e^{ix})\right\} \quad \Rightarrow \quad h'(\pi) = \text{Re}\left\{-ig'(-1)\right\} = 0.$$

Hence, since $h''(x) \geq 0$ for $x \in [0, \pi]$ and $h'(\pi) = 0$, then $h'(x) \leq 0$ for $x \in [0, \pi]$; this yields that $h$ is a monotonic non-increasing function over $[0, \pi]$. Now, we can apply Theorem C.2 to the $2\pi$-periodic function $h$.

Although $h(x)$ is convex, non-increasing for $x \in [0, \pi]$ and symmetric, we note that it may not satisfy the condition $h(x) \geq 0$ for $x \in [0, \pi]$. Revisiting the proof of Theorem C.2 shows that the condition $h(x) \geq 0$ over the interval $[0, \pi]$ can be replaced by the weaker condition $\int_0^\pi h(x)dx \geq 0$ which holds by assumption. This shows that $g_n \geq 0$ for all integers $n \geq 0$. □

## C.3 Applications: Non-Negativity Proofs for the Degree Distributions of Some Capacity-Achieving Ensembles over the BEC

In the field of codes on graphs, the validity of the construction of ensembles involves the verification that their degree distributions have non-negative positive power series expansions about $x = 0$.

### C.3.1 Non-Negativity Proof for the Ensemble of Check-Regular NSIRA Codes

In [8, Theorem 2], a capacity-achieving ensemble of non-systematic irregular repeat-accumulate (NSIRA) codes was introduced, achieving the capacity of a BEC under iterative message-passing decoding with bounded complexity per information bit. The d.d. of interest is given by Eq. (38), and the construction is only valid for a particular erasure rate, $p$, if $g(x) \triangleq \lambda(x)$ has a non-negative power series expansion for that $p$. Using Corollary C.1, we can define $h(x) = \text{Re}\left\{g\left(e^{ix}\right)\right\}$ and verify numerically, for any $p$, that $h''(x) \geq 0$ for $x \in [0, \pi]$, and also $\int_0^\pi h(x)\,dx \geq 0$. This approach appears to work for all $p \in [0, 1)$, hence it verifies the non-negativity of the power series expansion of the left d.d. $\lambda$. In [8, Appendix C], the non-negativity of the power series expansion of $\lambda$ was proved for $p \in [0, 0.95]$ where the proof there was quite involved. This approach extends the verification of the non-negativity of the power series expansion of $\lambda$ for all $0 \leq p < 1$, thus proving Conjecture 2 in [8, Section 3].

### C.3.2 Non-Negativity Proof for the Ensemble of Bit-Regular ARA Codes

Starting from (36), let

$$g(x) \triangleq \frac{R'(x)}{x}.$$

Since from (33), $\widetilde{\rho}^{-1}(0) = 0$, then $\widetilde{\rho}(0) = 0$, so the first non-negative coefficient of the power series expansion of $\widetilde{\rho}(x)$ about $x = 0$ is the one of $x$. Since $Q(x)$ in (35) is calculated by an appropriate scaling of the integral of $\widetilde{\rho}$ over the interval $[0, x]$ so that $Q(1) = 1$, then the first non-negative coefficient of the power series expansion of $Q(x)$ is the one of $x^2$. Finally, from (31), it follows that also the first non-negative coefficient of the power series expansion of $R(x)$ about $x = 0$ is the one of $x^2$. Hence, since $R(0) = R'(0) = 0$, then $R(x)$ has a power series expansion about $x = 0$ whose all coefficients are non-negative if and only if the function $g(x)$ possesses this property.

Using Corollary C.1, let us define $h(x) \triangleq \text{Re}\{g(e^{ix})\}$. It was verified numerically that if $p \leq 0.26$, then $h''(x) \geq 0$ for $x \in [0, \pi]$, and also $\int_0^\pi h(x)\,dx \geq 0$. This therefore proves that if $p \leq 0.26$, then the function $g(x)$ (and hence, also the d.d. $R(x)$) has a power series expansion about $x = 0$ whose all coefficients are non-negative.

### C.3.3 Non-Negativity Proof for the Ensemble of Self-Matched ARA Codes

In Section 6.1, we construct capacity-achieving ARA ensembles from self-matched LDPC codes. From (45) and (46), the left and right degree distributions can be rewritten in the form

$$\begin{aligned} L(x) &= \frac{K_1\big(-bx - \ln(1-bx)\big)}{1 + c_1\big(-bx - \ln(1-bx)\big)}, \\ R(x) &= \frac{K_2\big(-bx - \ln(1-bx)\big)}{1 + c_2\big(-bx - \ln(1-bx)\big)} \end{aligned} \tag{C.1}$$

where

$$\begin{aligned} K_1 &= -\frac{1}{p[b + \ln(1-b)]}, \quad K_2 = -\frac{1}{(1-p)[b + \ln(1-b)]}, \\ c_1 &= -\frac{1-p}{p[b + \ln(1-b)]}, \quad c_2 = -\frac{p}{(1-p)[b + \ln(1-b)]}. \end{aligned} \tag{C.2}$$

Note that $K_{1,2}$ and $c_{1,2}$ are positive for values of $p$ and $b$ in the range $0 < p < 1$ and $0 < b < 1$. Hence, in order to find the region of $p$ and $b$ where both degree distributions (i.e., $L(x)$ and $R(x)$) have non-negative power series expansion about $x = 0$, let us find conditions on the parameter $c$ which ensure that

$$g(x) = \frac{-x - \ln(1-x)}{1 + c\big(-x - \ln(1-x)\big)} \tag{C.3}$$

has a non-negative power series expansion about $x = 0$. The first few terms of the expansion are

$$\begin{aligned} g(x) &= \sum_{n=2}^{7} g_n(c) x^n \\ &= \frac{3x^2 + 2x^3}{6} + \frac{(1-c)x^4}{4} + \frac{(3-5c)x^5}{15} + \frac{(12 - 26c + 9c^2)x^6}{70} + \frac{(120 - 308c + 210)x^7}{840} + O\left(x^8\right). \end{aligned}$$

Applying Corollary C.1 directly to this function does not work because the function is neither non-increasing nor convex. Instead, we remove the first few terms of the power series and apply Corollary C.1 to

$$g(x) = \frac{1}{x^8}\left(\frac{-x - \log(1-x)}{1 + c\left[-x - \log(1-x)\right]} - \sum_{n=2}^{7} g_n(c) x^n\right).$$

Following the notation in Corollary C.1, let the function $h$ be

$$\begin{aligned} h(x) &\triangleq \text{Re}\left\{g(e^{ix})\right\} \\ &= \frac{1}{2}\Big[g(e^{ix}) + g(e^{-ix})\Big]. \end{aligned}$$

Numerically, we find that if $c \in [0, 0.6]$, then $h''(x) \geq 0$ for $x \in [0, \pi]$. Checking the first few terms by hand shows that $g_6(c)$ becomes negative first for $c > c^* \triangleq \frac{13 - \sqrt{61}}{9} \approx 0.5766$. Therefore, we find that the original power series expansion of the function $g$ in (C.3) is non-negative if and only if $c \in [0, c^*]$. From (C.1), it follows that the left and right d.d. ($L(x)$ and $R(x)$, respectively) have non-negative power series expansion about $x = 0$ if and only if

$$c_1 \leq \frac{13 - \sqrt{61}}{9}, \quad c_2 \leq \frac{13 - \sqrt{61}}{9}$$

which based on (C.2), is translated to the condition in (47).

Finally, we consider the question of whether the numerical verification of convexity on $[0, \pi]$ is reasonable. While one could also compute any finite number of terms in the power series expansion and verify their non-negativity, we note that the next term could always be negative. This new approach is different because, in many cases, the convexity can be verified numerically in finite time using complex interval arithmetic. The basic approach is subdividing the interval $[0, \pi]$ into a large number of small overlapping intervals. If interval output of the function $h$ evaluated on each subinterval of $[0, \pi]$ is non-negative, then this proves that $h$ is convex on $[0, \pi]$. If $h$ also depends on some parameter, then the parameter interval can also be subdivided into small overlapping subintervals. If the first test succeeds for each parameter subinterval, then we have shown that $h$ is convex on $[0, \pi]$ for all parameter values as well.

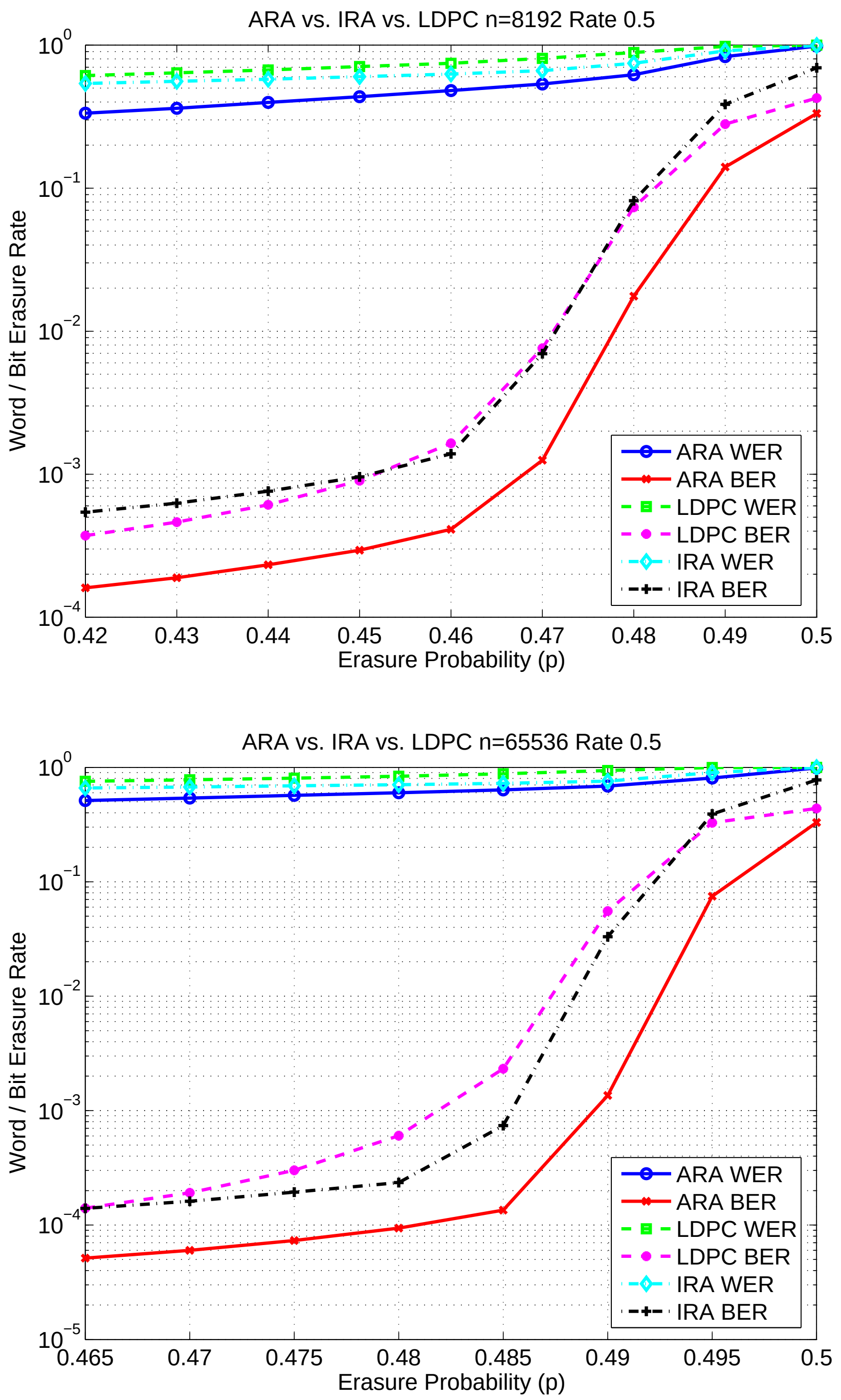

Figure 8: Simulations for the ensembles of ARA and NSIRA codes constructed from LDPC codes with self-dual degree distributions (see Sections 6.1 and 6.3), and the ensemble of right-regular LDPC codes in [2]. The plots refer to block lengths of 8192 and 65536 bits (see upper and lower plots, respectively) and a design rate of 0.5 bits per channel use. No high-rate outer code is assumed.

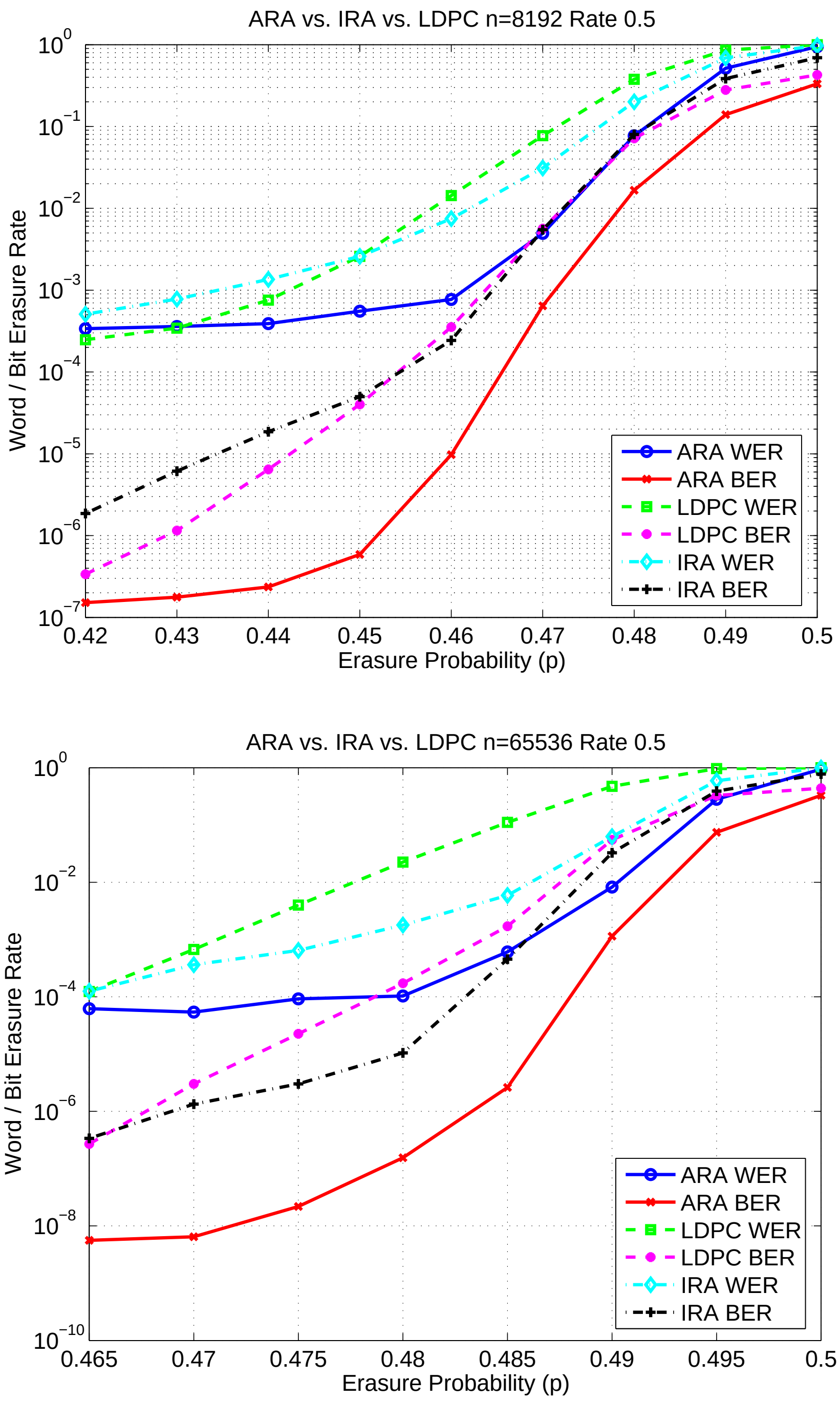

Figure 9: Simulations for the ensembles of ARA and NSIRA codes constructed from LDPC codes (see Sections 6.1 and 6.3), and right-regular LDPC codes [2]. The plots refer to block lengths of 8192 and 65536 bits (see upper and lower plots, respectively) and a design rate of 0.5 bits per channel use. Since the ensemble averaged performance is simulated, high-rate outer codes (rates $\frac{8179}{8192}$ and $\frac{65520}{65536}$, respectively) are used to lower the error floor due to small stopping sets. These outer codes are chosen uniformly at random from the ensemble of the binary linear block codes and their rate loss is neglected.

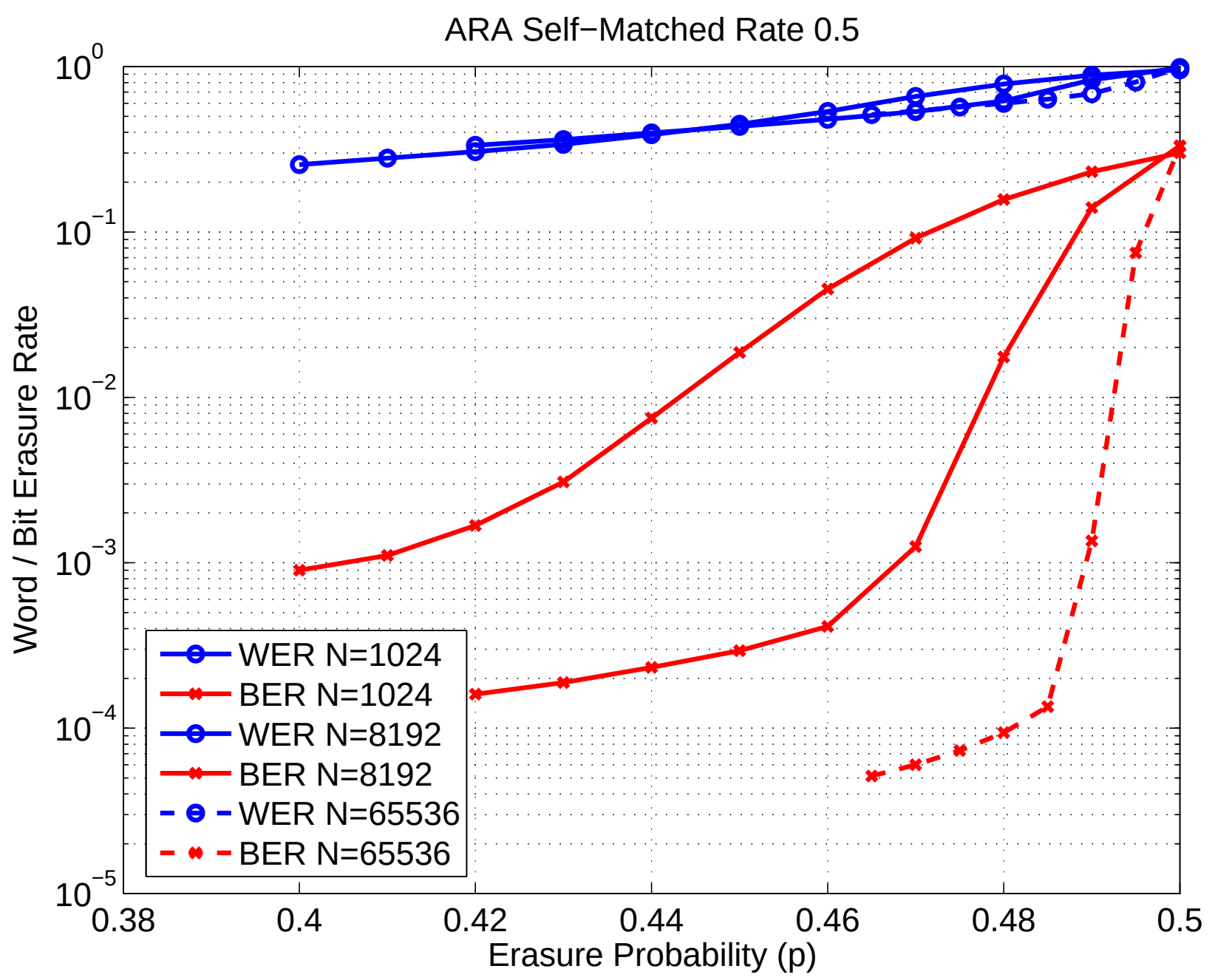

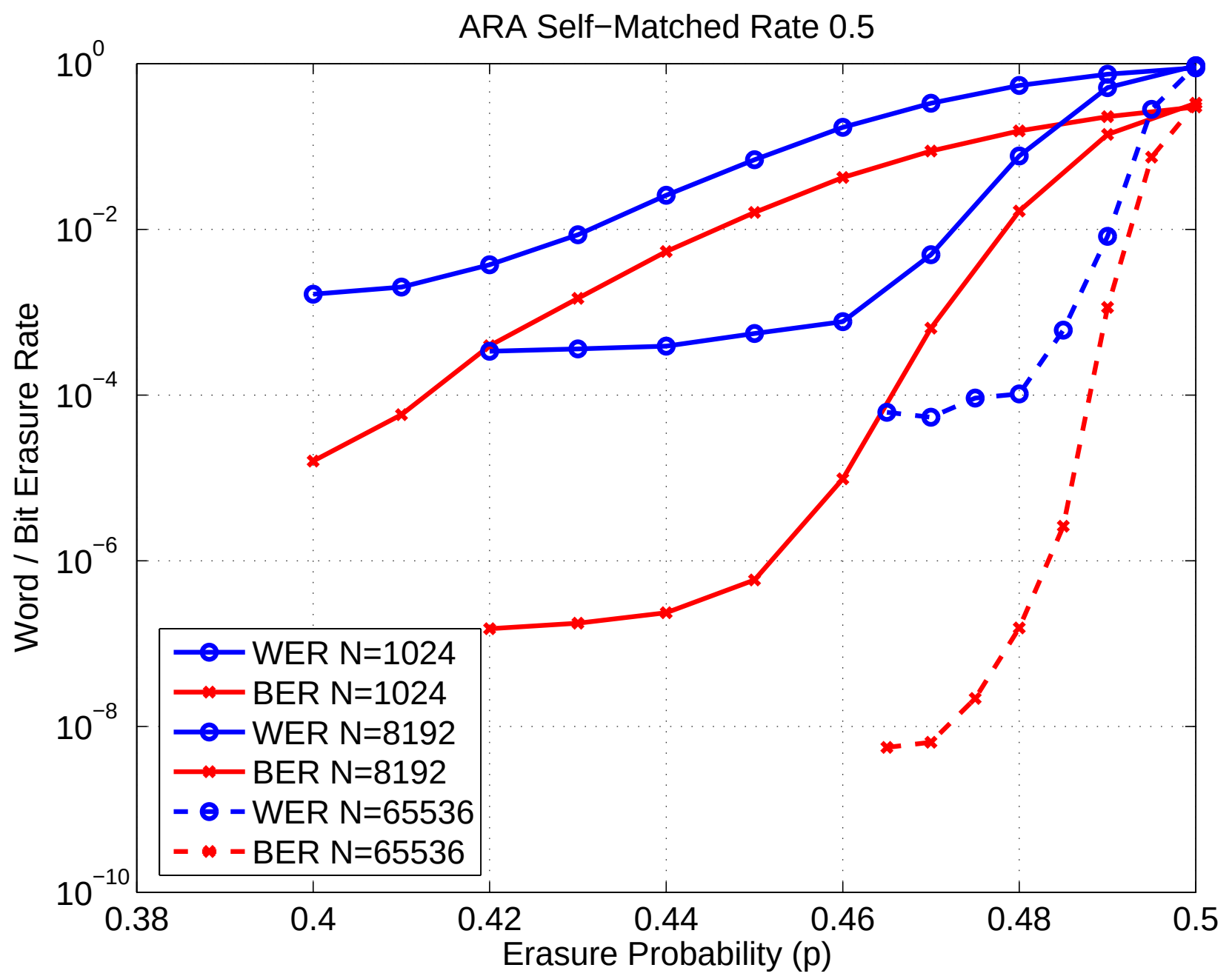

Figure 10: Simulations for the ensemble of ARA codes in Section 6.1, constructed from the ensemble of LDPC codes whose degree distributions are matched to themselves. The design rate of the ensemble is 0.5 bits per channel use. The upper plot refers to the case where there is no outer high-rate code, and the lower plot refers to the case where there is such a code.

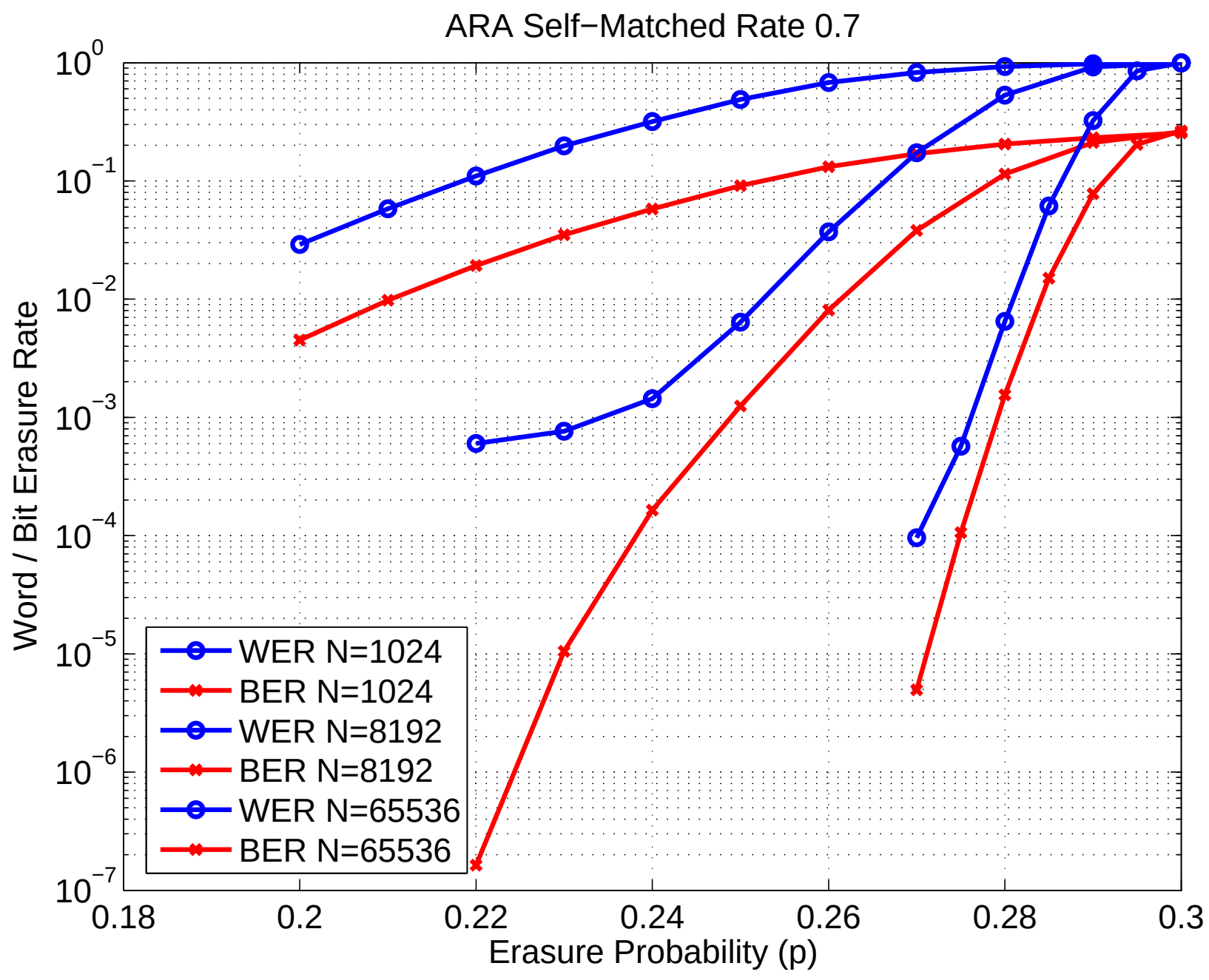

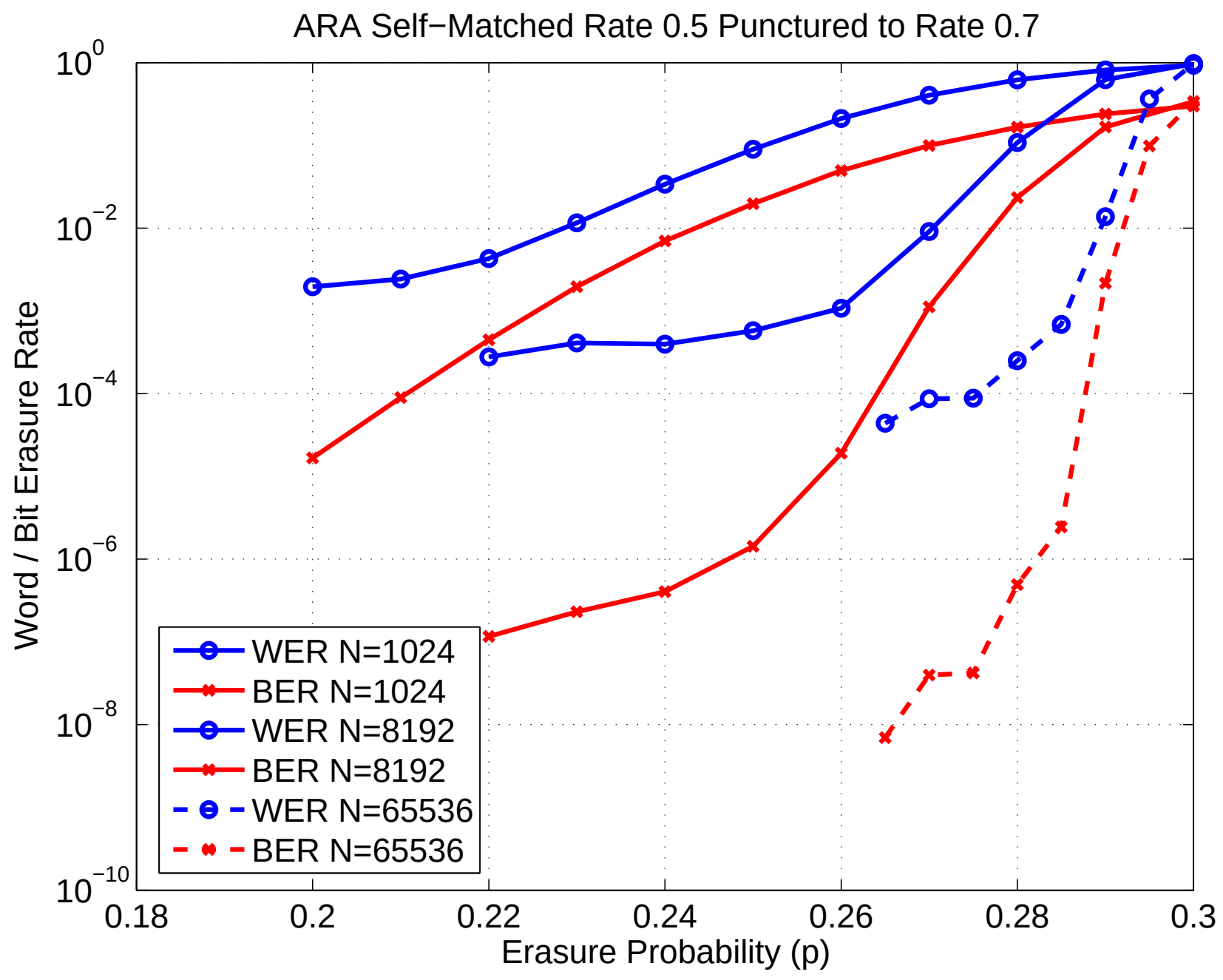

Figure 11: Simulations for the ensemble of self-matched ARA codes whose rate is 0.7 bits per channel use, having high-rate outer codes (the rate of the outer code is $\frac{1014}{1024}$, $\frac{8179}{8192}$ and $\frac{65520}{65536}$, for a block length of 1024, 8192 and 65536 bits, respectively.) The upper plot refers to the case where the ensemble of self-matched ARA codes is directly designed for a rate of 0.7 (without puncturing), and the lower plot refers to the design of the self-matched ARA ensemble for a rate of 0.5, and then increasing the rate to 0.6 by random puncturing of the code bits.

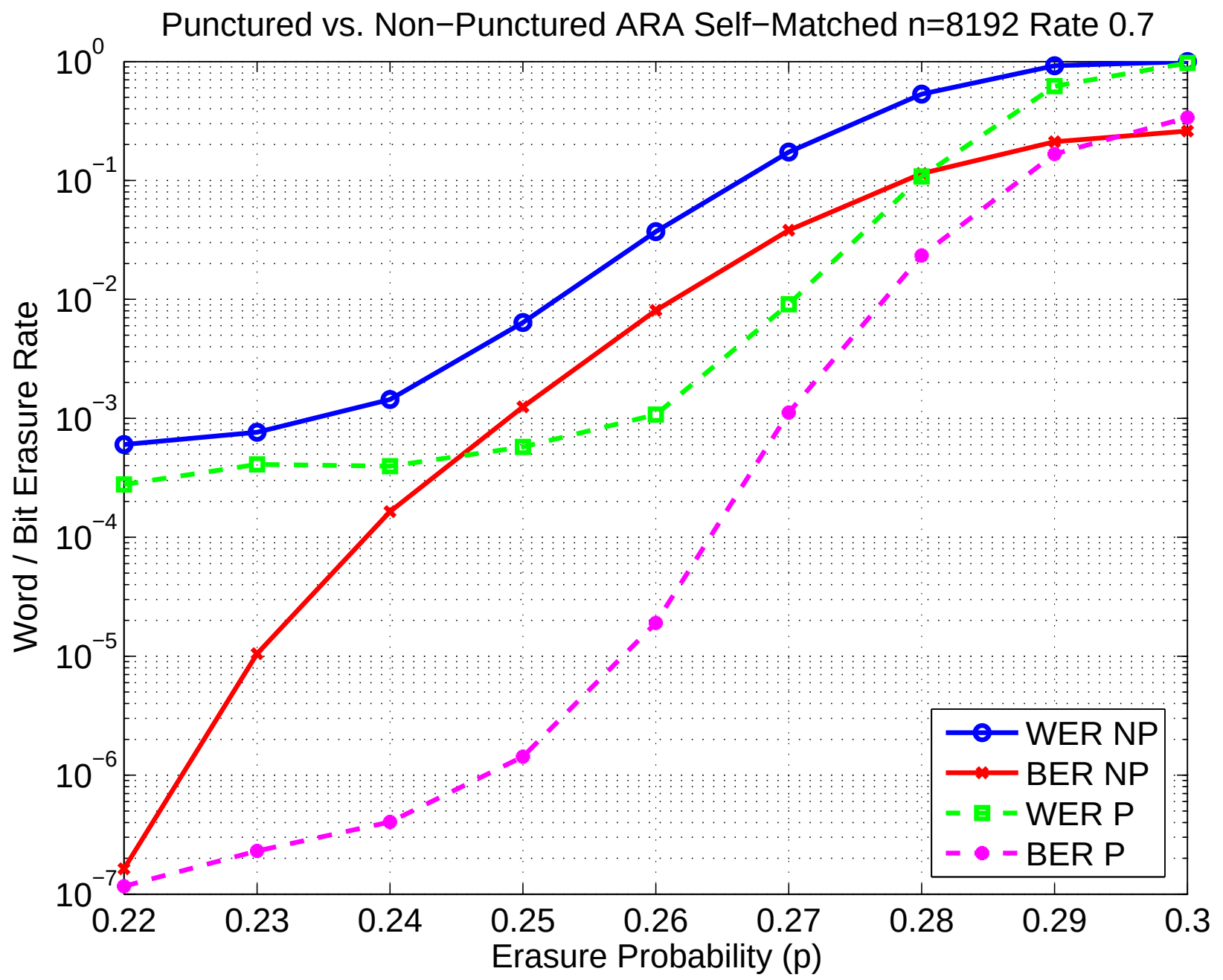

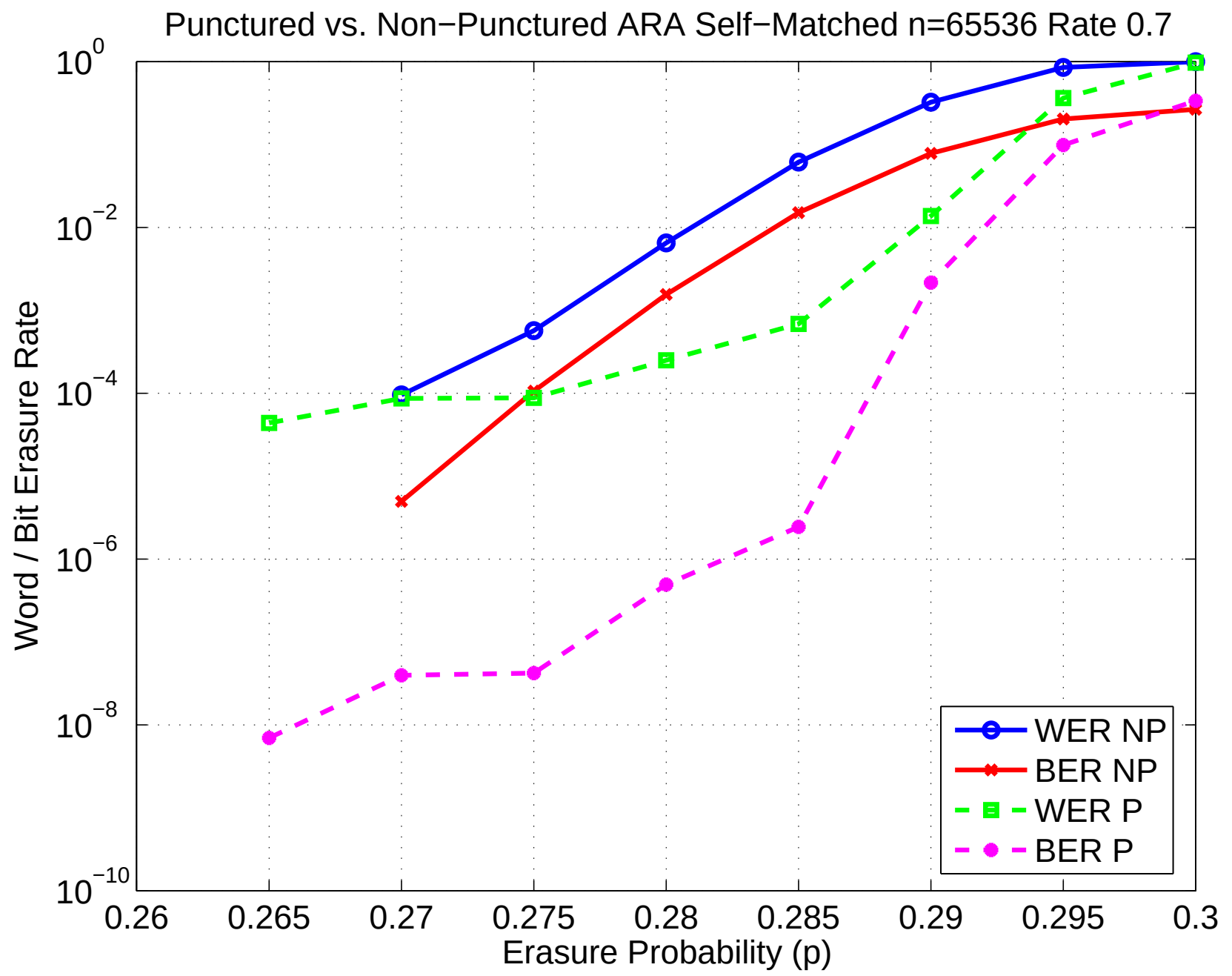

Figure 12: Simulations for the ensemble of punctured ARA codes in Section 6.1 where we compare the case where the self-matched ensemble is designed directly to a rate of 0.7 bits per channel use, versus the case of designing the ensemble for a rate of 0.5 and increasing the rate by puncturing. The upper and lower plots refer to block lengths of 8192 and 65536 bits, respectively. P and NP stand for 'punctured' and 'non-punctured'.

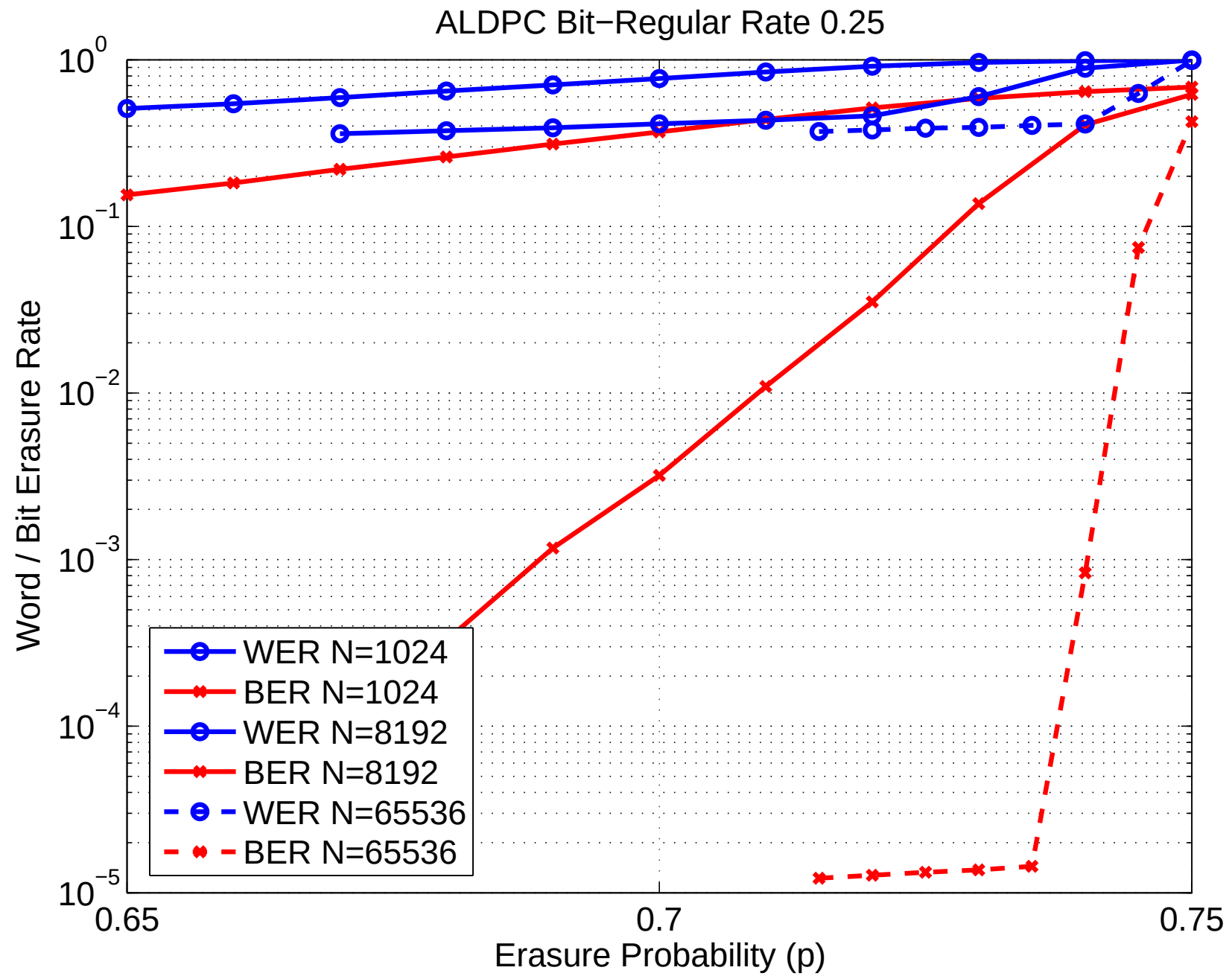

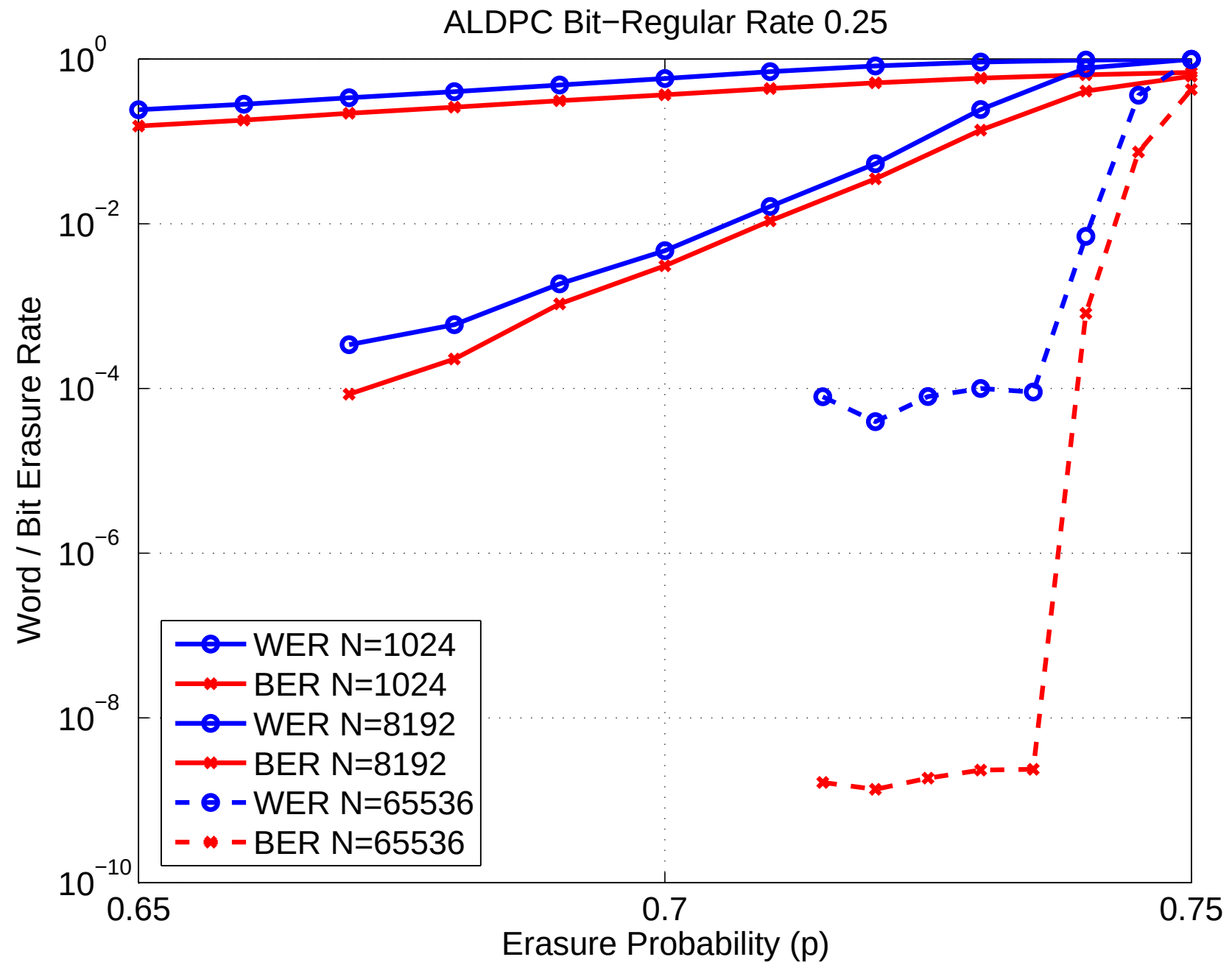

Figure 13: Simulations for the ALDPC-BR3 ensemble (i.e., bit-regular accumulate-LDPC ensemble whose bit nodes have degree-3); the design rate of this ensemble is set to $\frac{1}{4}$ bits per channel use. The upper plot refers to the case where there is no high-rate outer code, and the lower plot refers to the case where there is such an outer code. The plots refer to block lengths of 1024, 8192 and 65536 bits; for the lower plot and these block lengths, the rate of the outer random code is equal to $\frac{1014}{1024}$, $\frac{8179}{8192}$ and $\frac{65520}{65536}$, respectively.